\documentclass[journal]{IEEEtran}
\usepackage{cite}

\usepackage{amsmath,amssymb,amsfonts}
\usepackage{algorithmic}
\usepackage{gensymb}
\usepackage{amsthm}

\usepackage{graphicx,subfigure}
\usepackage{textcomp}
\usepackage{xcolor}

\usepackage{subfigure}
\usepackage{pdfpages}
\usepackage{diagbox}
\usepackage{colortbl}
\definecolor{mygray}{gray}{0.85}
\usepackage{array}
\usepackage{caption}
\usepackage[linesnumbered,boxed,ruled,commentsnumbered,noend]{algorithm2e}

\usepackage{eqparbox}

\usepackage{bbding}
\usepackage{tabularx}
\usepackage{tabularray}
\usepackage{booktabs}
\usepackage{amssymb}
\usepackage{xcolor}
\usepackage{array}
\ifCLASSINFOpdf
\else
\fi

\begin{document}
%
\title{Cross-Problem Solving for Network Optimization: Is Problem-Aware Learning the Key?}

\author{Ruihuai~Liang,~\IEEEmembership{Student Member,~IEEE,} Bo~Yang,~\IEEEmembership{Senior Member,~IEEE,} Pengyu~Chen, Xuelin Cao,\\ \IEEEmembership{Senior Member,~IEEE,} Zhiwen Yu,~\IEEEmembership{Senior Member,~IEEE,}  
H. Vincent Poor,~\IEEEmembership{Life Fellow,~IEEE}, and Chau~Yuen,~\IEEEmembership{Fellow,~IEEE}  
\thanks{

This work was supported in part by the National Natural Science Fund for Excellent Young Scientists Fund Program (Overseas), in part by the Qin Chuang Yuan Fund Program under Grant QCYRCXM-2022-358, in part by ``the Fundamental Research Funds for the Central Universities" under Grant No.  D5000250296, and in part by Key Research and Development Program of Shaanxi (Program No. 2025GH-YBXM-049). The work of X. Cao was supported in part by the Qin Chuang Yuan Fund Program under Grant QCYRCXM-2022-240. The work of Z. Yu was supported in part by the National Fund for Distinguished Young Scholars (No. 62025205). The work of H. V. Poor was supported in part by an Innovation Grant from Princeton NextG.

R. Liang, B. Yang and P. Chen are with the School of Computer Science, Northwestern Polytechnical University, Xi'an, Shaanxi, 710129, China (email: liangruihuai$\_$npu, cpy$\_$npu@mail.nwpu.edu.cn, yang$\_$bo@nwpu.edu.cn). 

Z. Yu is with the School of Computer Science, Northwestern Polytechnical University, Xi'an, Shaanxi, 710129, China, and Harbin Engineering University, Harbin, Heilongjiang, 150001, China (email: zhiwenyu@nwpu.edu.cn).

X. Cao is with the School of Cyber Engineering, Xidian University, Xi'an, Shaanxi, 710071, China (email: caoxuelin@xidian.edu.cn). 


H. V. Poor is with the Department of Electrical and Computer Engineering, Princeton University, Princeton, NJ 08544, USA (email: poor@princeton.edu).

C. Yuen is with the School of Electrical and Electronics Engineering, Nanyang Technological University, Singapore (email: chau.yuen@ntu.edu.sg).

Corresponding author: Bo Yang.
}
}

\markboth{Journal of \LaTeX\ Class Files,~Vol.~XX, No.~XX, April~2025}%
{Shell \MakeLowercase{\textit{et al.}}: A Sample Article Using IEEEtran.cls for IEEE Journals}


\maketitle

\begin{abstract}
As intelligent network services continue to diversify, ensuring efficient and adaptive resource allocation in edge networks has become increasingly critical. Yet the wide functional variations across services often give rise to new and unforeseen optimization problems, rendering traditional manual modeling and solver design both time-consuming and inflexible. This limitation reveals a key gap between current methods and human solving — the inability to recognize and understand problem characteristics. It raises the question of whether problem-aware learning can bridge this gap and support effective cross-problem generalization. To answer this question, we propose a problem-aware diffusion (PAD) model, which leverages a problem-aware learning framework to enable cross-problem generalization. By explicitly encoding the mathematical formulations of optimization problems into token-level embeddings, PAD empowers the model to understand and adapt to problem structures. {Extensive experiments across ten representative network optimization problems show that PAD generalizes well to unseen problems while avoiding the inefficiency of building new solvers from scratch, yet still delivering competitive solution quality.} Meanwhile, an auxiliary constraint-aware module is designed to enforce solution validity further. The experiments indicate that problem-aware learning opens a promising direction toward general-purpose solvers for intelligent network operation and resource management. Our code is open source at https://github.com/qiyu3816/PAD.
\end{abstract}

\begin{IEEEkeywords}
Network optimization, resource allocation, problem-aware learning, generative diffusion model.
\end{IEEEkeywords}

\section{Introduction}
\IEEEPARstart{T}{he} vision of future wireless communication systems has been widely built in both academia and industry \cite{roadmap6G,connected2collective,cao2024aiPROIEEE,hermes2024autonomous,oRAN2024Evolution6G}, with fundamental AI models capable of general problem-solving being pursued as a revolutionary technology. These intelligent fundamental AI models are expected to enable autonomous and general network operations across various tasks, such as network resource management \cite{gdsg2024liang} and data analysis \cite{dataAnalysisHotnet}. Notably, the continuous emergence of novel intelligent services in wireless networks is driving an unprecedented demand for efficient allocation of network resources like power and spectrum. In future wireless networks, different types of tasks—such as power allocation and spectrum allocation—will coexist within edge networks, while edge networks themselves may differ significantly in functionality and environmental settings (e.g., computation offloading networks and edge sensing networks), which inherently raises diverse network optimization problems with varying objectives, constraints, and mathematical formulations \cite{oRAN2024Evolution6G,subnetwork2023X,userCentric2024debbah,otter2023mobicom,shi2023optimizationIn6G,yang2025frontiersgenerativeainetwork}. However, most existing network optimization approaches treat each problem in isolation, failing to address them in a unified and seamless manner. Consequently, cross-problem network optimization is a key capability for the intelligent fundamental models of next-generation wireless communication systems.

Cross-problem network optimization presents several challenges with existing approaches. First, most methods \cite{tango2023infocom,risCov2022infocom,gnn2023jsac,gdsg2024liang,gdmRL2024jsac,gdmContract2025tmc} rely on manually analyzing predefined problems and designing corresponding algorithms or models. This approach is labor-intensive thus impractical for achieving cross-problem generalization at scale. Additionally, AI philosophies theoretically capable of general-purpose problem-solving \cite{fujimoto2025generalpurposeRL} often require extensive customization of hyperparameters and algorithms in practice. For example, problem-specific value networks \cite{d2sacTMC,feat2023infocom} introduce inefficiencies that make them unsuitable for cross-problem network optimization. Finally, AI-based network optimization methods \cite{agi2023incentive,gdplan2025ton,gnn2023jsac} have long been criticized for their inability to strictly enforce hard constraints. These methods often rely entirely on externally curated rules to satisfy constraints, limiting their applicability. 

Overall, most existing network optimization approaches either depend on expert-driven designs, making them inefficient, or function as statistical black boxes, lacking problem awareness, as shown in Fig. \ref{fig_outline}. They fail to leverage the model itself to learn to recognize and understand problem characteristics (e.g., mathematical structures) for problem-solving, which is how humans approach cross-problem tasks. 

In this context, a key question raised: \textit{Can problem-aware learning enable models to develop a meaningful understanding of problem objectives and achieve cross-problem generalization?} Inspired by recent advances in using large language models (LLMs) for mathematical reasoning \cite{AI4MathFrontier}, math question similarity calculation \cite{neuralSolver2022PNAS}, and optimization problem modeling \cite{telecomGPT}, we argue that the mathematical formulations of network optimization problems also contain instructive information that can help models better understand problem characteristics. Incorporating these representations into the model’s inference process can enhance the capacity for cross-problem solving. {To address the aforementioned challenges, we propose PAD, a problem-aware diffusion model designed for cross-problem solution generation in network optimization within a single fundamental model.} PAD explores a novel problem-aware approach that integrates the rich mathematical representations of optimization problems into model learning. To achieve this, PAD leverages feature embeddings of network optimization problem formulations from existing mathematical LLMs during both training and solution generation. 

Specifically, we extract the most informative features from the raw token embeddings of MathBERT \cite{mathbert} or DeepSeekMath \cite{deepseek2024math} using a pooling mechanism based on token similarity scores. PAD adopts the latent diffusion model (LDM) \cite{StableDiff2022CVPR} with a Transformer-based backbone \cite{attention2017transformer}, embedding both input parameters and optimization solutions into a latent space through a pre-trained encoder-decoder architecture. By conditioning the denoising process on both the embedded input parameters and the pooled problem representations, PAD enables explicit recognition and learning of mathematical structures, granting it a degree of generalization to entirely unseen optimization problems. To further enhance constraint satisfaction in the generated solutions, PAD introduces a constraint-aware module, which is used only during training. This module approximates the original constraints as complex non-linear functions and introduces a differentiable constraint loss into the PAD's objective. It takes the problem representations, input parameters, and current solution as input, and outputs a binary classification signal (0-1) indicating constraint satisfaction. By integrating this validation loss, PAD strengthens adherence to hard constraints in network optimization, improving solution feasibility and quality across diverse problem settings.

Our main contributions lie in providing exploratory answers to the following questions:\hspace{-8mm}
\begin{itemize}
    \item \textbf{Can problem-aware learning enhance cross-problem generalization in network optimization?} We propose PAD that achieves problem-aware generalization within a single foundational model. By incorporating mathematical feature embeddings of optimization problems, PAD demonstrates the value of learning the mathematical structures of network optimization problems, offering a novel approach for models to grasp problem characteristics and achieve cross-problem solving.
    \item \textbf{How can raw problem-specific feature embeddings be effectively utilized in problem-aware learning?} To effectively extract key information from the token sequence of the raw problem embedding, we design a rank pool mechanism based on intra-sequence token similarity scores. This approach not only extracts essential information from the raw embedding and remains sensitive to structural variations, but also provides greater potential to capture the most distinctive features that differentiate one problem from others. Empirically, our method outperforms other standard pooling techniques.
    \item \textbf{Can a constraint-aware module improve the feasibility of generated solutions?} PAD introduces a constraint-aware module, which markedly improves constraint satisfaction by incorporating a constraint loss during training. This provides an efficient mechanism for ensuring that the model's outputs comply with the hard constraints of the original optimization problems.
\end{itemize}

\section{Preliminaries}\label{sec_preliminaries}
\subsection{Solution Generation for Network Optimization}
Solution generation methods based on LLMs \cite{llm2024sigcomm,userCentric2024debbah,nonConvex2025llm} and generative diffusion models (GDMs) \cite{gdsg2024liang,gdplan2025ton,wang2024DFSS,liang2025diffsg,liang2025iotj} have achieved remarkable success in network optimization tasks. These approaches leverage the strong logical reasoning capabilities of LLMs and the powerful solution distribution modeling of diffusion models to produce superior solutions for specific problems. As data-driven approaches, generative models differ fundamentally from discriminative models. While discriminative models learn conditional probability distributions, generative models learn joint distributions, which makes their inference process inherently a form of sampling rather than classification or regression. This sampling-based inference enables parallel generation and provides a more comprehensive understanding of the task compared to the fixed-output nature of conventional regression models. Generative AI (GenAI)-based frameworks for network management and optimization are rapidly gaining momentum in the communication network community. With their strong scaling laws, GenAI is increasingly seen as a promising candidate for serving as a foundational model in future communication systems.

\begin{figure}[b]
\captionsetup{font={small}}
\centerline{\includegraphics[width=3.5in]{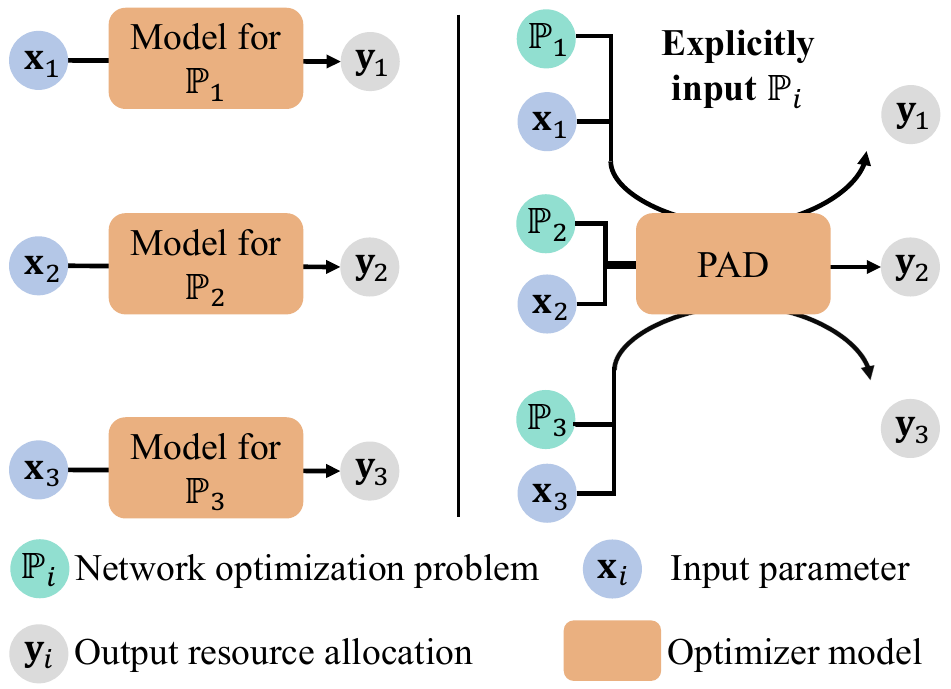}}
\caption{ Comparison of PAD with other isolated methods for cross-problem network optimization. PAD trains once and adapts to new problems by explicitly inputting $\mathbb{P}_i$ from formulation embedding, whereas traditional methods require re-training, even redesigning, for each individual problem.}
\label{fig_outline}
\vspace{-0.38cm}
\end{figure}

\subsection{Cross-Problem Solving}
Existing approaches have explored cross-problem generalization from various perspectives. \cite{crossProblemCO2025} analyzes the feasibility of joint training on different optimization problems within a single model by computing the cosine similarity between their gradient descent directions. \cite{lei2025boostinggeneralization} enhances cross-problem and cross-scale solution generation by conditioning a pre-trained model on energy-guided scores derived from problems similar to the original one. Meanwhile, \cite{llm2024sigcomm} and \cite{d2sacTMC} adopt LLMs and GDMs, respectively, as actors, leveraging deep reinforcement learning (DRL) paradigms for cross-problem solving. However, these methods often rely on prior knowledge such as the optimal solution score \cite{lei2025boostinggeneralization} or a fixed problem space \cite{llm2024sigcomm}, or require building additional modules from scratch for each new task \cite{crossProblemCO2025,d2sacTMC}. In practical network optimization scenarios, neither the optimal objective score nor the full scope of new problems is typically available, and retraining models from scratch for every new case is infeasible.

\subsection{Problem-Aware Learning}
Problem-aware learning refers to a class of learning methods being explicitly informed or guided by the structure, objectives, or constraints of the underlying problem being solved, instead of learning patterns purely from data in an unconscious way \cite{formalMethod2022AsProblemAware}. For example, \cite{deepmind2024LLMasOptimizer} employs LLMs to iteratively generate solutions for simple optimization problems based on natural language descriptions, and \cite{nonConvex2025llm} leverages domain-specific LLMs to produce resource allocation solutions for network optimization problems described in natural language. In addition, \cite{HotNets2023FormalMethodIntegration} advocates integrating formal methods (such as mathematical algorithms or rules) with machine learning, similar to problem-aware learning. However, \cite{HotNets2023FormalMethodIntegration} is limited to combining simple constrained problems into losses in the form of Lagrange functions, and is challenging to apply to network optimization problems where non-differentiable functions are frequently encountered.

Other related works explicitly leverage mathematical expressions to enhance the capabilities of LLMs on various tasks. In \cite{telecomGPT}, the domain-specific LLM ``TelecomGPT" is used to generate mathematical formulations from natural language descriptions. These generated formulations are then compared with ground truth using MathBERT and cosine similarity, demonstrating that LLMs can extract meaningful representations from network optimization problems. In the AI for Math community \cite{AI4MathFrontier,neuralSolver2022PNAS}, reasoning based on mathematical descriptions and problem formulations has been deeply investigated. Compared to relying on optimal solution scores or predefined problem spaces, acquiring mathematical formulations of network optimization problems is more realistic and informative. It can substantially benefit the design of problem-aware learning.

However, the natural language process (NLP)-centric framework of LLMs lacks precision in solution generation, often requiring repeated iterations based on feedback, and their training and inference costs are disproportionately high relative to the task. Although capable of performing direct problem-aware learning, they are not suitable for real-time implementation in network optimization.

At the same time, the potential of GDMs for cross-problem generalization has been validated in the AI community \cite{he2023haoran,fan2025taskagnostic}. Nevertheless, these works primarily focus on automatic control problems, which differ significantly from network optimization in both formulation and application. Therefore, we believe that integrating GDMs with problem-aware learning is a promising direction for achieving cross-problem solving in network optimization.

\section{System Model}

\begin{figure*}[t]
\centering
\captionsetup{font={small}}
\centerline{\includegraphics[width=6.25in]{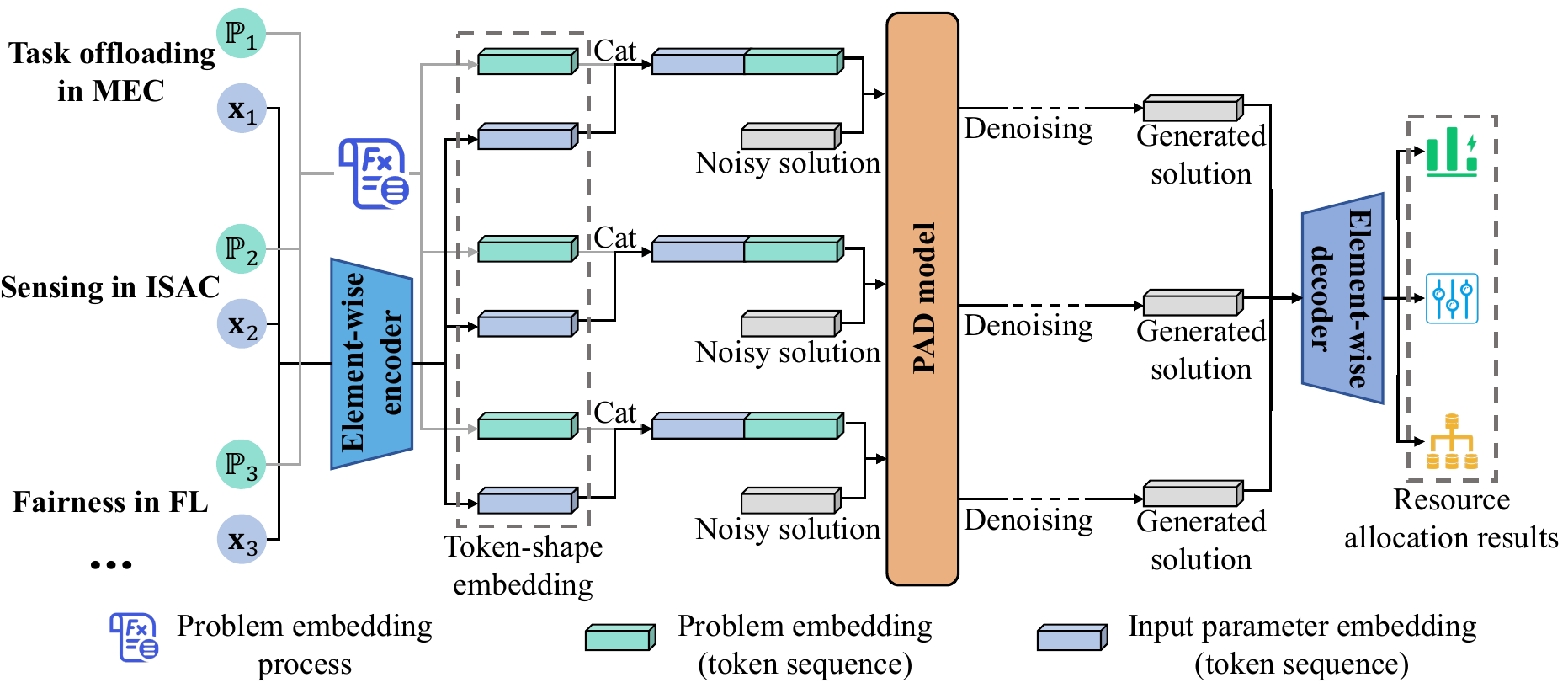}}
\caption{Implementation framework of PAD.}
\label{fig_framework}
\vspace{-0.38cm}
\end{figure*}

\subsection{Cross-Problem Network Optimization}

Network optimization problems arise from the need to allocate limited resources, such as communication and computation capacity, among various tasks. The objective is to determine the optimal resource allocation strategy that maximizes service performance under resource constraints, which are typically formulated and solved as mathematical optimization problems \cite{shi2023optimizationIn6G}. These mathematical formulations usually have the following form
\begin{equation}\label{eq_stantard_optimization}
    \mathbb{P}:\ \min_{\mathbf{y}}f(\mathbf{x},\mathbf{y}),\ \rm{s.t.}\ g(\mathbf{x},\mathbf{y})>0,
\end{equation}
where $M$-dimensional vector $\mathbf{x}\in\mathbb{R}^M$ denotes the input parameters, such as network states and task-related information, while $N$-dimensional vector $\mathbf{y}\in\mathbb{R}^N$ represents the optimization variables, such as power allocation and spectrum assignment. The functions $f:\mathbb{R}^{M+N}\rightarrow\mathbb{R}$ and $g:\mathbb{R}^{M+N}\rightarrow\mathbb{R}$ correspond to the objective function and the constraint function, respectively. {In network optimization, $f$ typically corresponds to global objectives that need to be maximized or minimized, such as transmission rate, throughput, or resource cost. $g$, on the other hand, is usually used to enforce physical resource constraints and prevent logical conflicts.}

Cross-problem network optimization refers to the ability of a method to generalize across structurally and semantically different optimization problems, while preserving solution quality and not requiring task-specific redesign. The problems in this set may differ in terms of objective function structure, constraint form, and the semantic interpretation of inputs and outputs. As illustrated in Fig. \ref{fig_outline}, different network optimization problems such as $\mathbb{P}_1$ and $\mathbb{P}_2$ may have distinct mathematical semantics in their objective functions \( f_1 \) and \( f_2 \), as well as in their constraint functions \( g_1 \) and \( g_2 \). They may also differ in their input parameters \( \mathbf{x}_1, \mathbf{x}_2 \) and output variables \( \mathbf{y}_1, \mathbf{y}_2 \). 

Traditional isolated optimizers rely heavily on intensive human effort, limiting their applicability to narrowly scoped scenarios. Cross-problem network optimization presents a promising approach to efficient, automated, and scalable network operations in increasingly dynamic environments. However, as described in Section \ref{sec_preliminaries}, these models do not explicitly incorporate the mathematical structure of the problems during training, resulting in limited adaptability to new tasks. 

We aim to enhance cross-problem generalization by embedding problem-specific mathematical information into model training, enabling direct adaptation to previously unseen tasks. These problem feature embeddings can be efficiently extracted from existing math-domain LLMs, and their informativeness and mathematical value have already been well recognized and validated \cite{AI4MathFrontier,neuralSolver2022PNAS,telecomGPT}. Although perfect generalization across all network optimization problems may not be achievable, our method aims to generalize over a broad problem space without requiring a fixed number of predefined tasks.

\subsection{Selected Optimization Problems}\label{sec_selected_problems}
In this section, we provide an overview of the set of network optimization problems considered in our implementation. The problem selection is based on three key principles: 1) the problems must involve the allocation of major network resources; 2) the problems should include diverse types of input features and optimization variables with different physical meanings; 3) the problems should cover a wide range of objectives and constraints to reflect the diversity of optimization problems encountered in real-world scenarios. Specifically, we consider the following ten problems:
\begin{enumerate}
    \item Power allocation aiming to maximize total transmission rate, subject to minimum rate constraints per link;
    \item Spectrum resource block allocation aiming to maximize total transmission rate, subject to minimum rate constraints per link;
    \item Spectrum resource block allocation aiming to maximize total transmission rate, subject to fairness constraints;
    \item Spectrum resource block allocation aiming to maximize spectral efficiency, subject to minimum rate constraints per link;
    \item Power allocation aiming to maximize power efficiency, subject to minimum rate constraints per link;
    \item {Cache refresh rate optimization aiming to minimize expected hit latency;}
    \item {Joint spectrum and power allocation aiming to maximize total transmission rate, subject to minimum rate constraints per link;}
    \item {Cache decision optimization aiming to minimize expected hit latency;}
    \item {Joint offloading decision and computational resource optimization aiming to minimize overall task offloading cost;}
    \item Power allocation aiming to minimize the maximum transmission delay.
\end{enumerate}

The detailed mathematical formulations of these problems are provided in the \textbf{Appendix}.

{These problems include several common resource types: power \cite{nonConvex2025llm,powerAlloc2022aerial,powerAlloc2023uav}, spectrum \cite{wang2024DFSS,spectrumSharing2019,dynamicSpectrumAlloc2023,joint2021twc}, caching resources \cite{refresh2019optimize,caching2017jsac,content2018caching}, and computational resources \cite{yang2021MTFNN,xue2025jointtaskoffloadingresource}, which together represent the core decision variables in modern network optimization.} The selected problems involve both continuous (real-valued) and discrete (integer-valued) variables, reflecting the real-world diversity of decision spaces. Standard parameters, such as channel gains, data volumes, and noise power, are incorporated to further enhance the realism. {The objectives and constraints of these problems are inspired by key application scenarios, including multi-access edge computing (MEC) networks \cite{mec2017mao,yang2021MTFNN}, integrated sensing and communication (ISAC) networks \cite{cao2024aiPROIEEE,mmWave2018comst}, federated learning (FL) systems \cite{survey2023federated}, and content caching and delivery (CDN) networks \cite{caching2017jsac,content2018caching}.} 

{The above problems cover the common mathematical optimization types encountered in network optimization. Specifically, the first nine problems are grouped into three clusters, each containing three problems and corresponding to a representative optimization paradigm: convex optimization, non-convex optimization, and mixed-integer or combinatorial optimization. The final problem, power allocation to minimize the maximum transmission delay, is formulated as a bi-level optimization problem and is inherently more challenging; therefore, it is presented separately. In the following model design, only one representative problem from each cluster is used for training, while the remaining problems are reserved for testing and evaluation of generalization.} By evaluating the proposed PAD across this selected and diverse problem set, we aim to demonstrate its effectiveness in handling different optimization structures and generalizing across diverse tasks.

\section{Methodology}

\begin{figure*}[t]
\centering
\centerline{\includegraphics[width=5.45in]{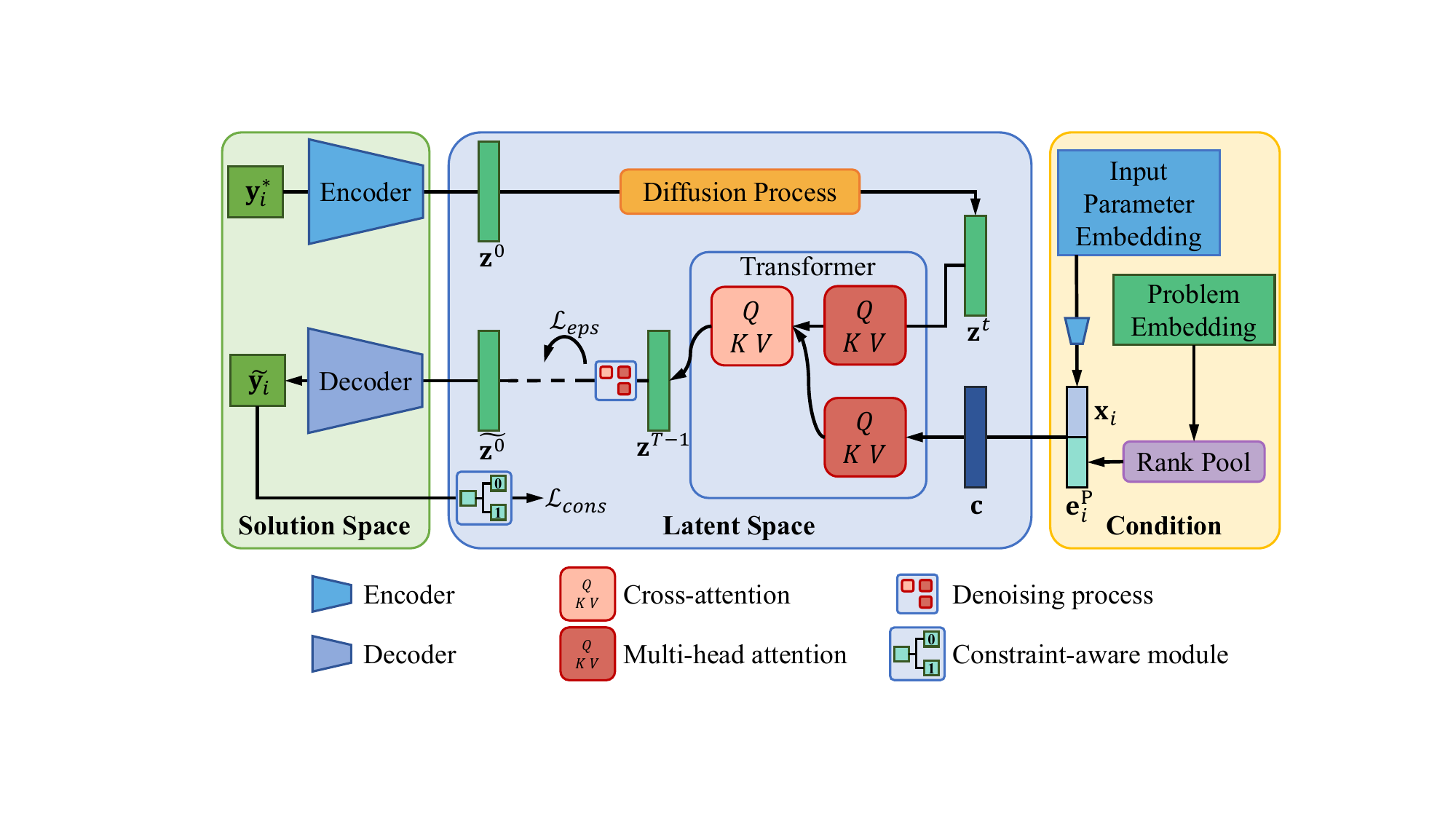}}
\caption{Neural network structure of PAD.}
\label{fig_neural}
\vspace{-0.38cm}
\end{figure*}

\subsection{Basics}\label{sec_theory_basics}
To enable problem-aware learning, we first illustrate the theoretical advantages of incorporating mathematical problem embedding into the model by analyzing its impact on the learning objective. Let the process of problem feature embedding be denoted as $\mathbf{e}_i\!=\!\mathrm{PE}(f_i,g_i)$, where $i$ represents the index of a problem instance in the problem set, and $\mathbf{e}_i$ is the resulting embedding sequence. Without incorporating problem embeddings, the theoretical learning objective becomes 
\begin{equation}\label{eq_no_prob_emb_tgt}
    p(\mathbf{y}_i,\mathbf{e}_i|\mathbf{x}_i)=\frac{p(\mathbf{y}_i,\mathbf{e}_i,\mathbf{x}_i)}{p(\mathbf{x}_i)},
\end{equation}
where the model must learn a joint mapping from the input parameter $\mathbf{x}_i$ to both the associated problem $\mathbf{e}_i$ and its corresponding solution.

Similarly, the learning objective after incorporating problem embeddings becomes 
{\begin{equation}\label{eq_with_prob_emb_tgt}
    p(\mathbf{y}_i|\mathbf{e}_i,\mathbf{x}_i)=\frac{p(\mathbf{y}_i,\mathbf{e}_i,\mathbf{x}_i)}{p(\mathbf{e}_i,\mathbf{x}_i)}.
\end{equation}}

{In the context of network optimization, different problems may share the same network state parameters and resource allocation variables, yet certain specialized parameters and variables may correspond exclusively to specific problems. As a result, the input parameter \( \mathbf{x} \) and the problem embedding \( \mathbf{e} \) are not conditionally independent, and their correspondence is inherently non-deterministic. Therefore, we have 
\begin{equation}\label{eq_conditional_e_x}
    p(\mathbf{e}_i,\mathbf{x}_i)=p(\mathbf{e}_i|\mathbf{x}_i)p(\mathbf{x}_i)\leq p(\mathbf{x}_i).
\end{equation}
}

Combining Eq. (\ref{eq_no_prob_emb_tgt}), (\ref{eq_with_prob_emb_tgt}), (\ref{eq_conditional_e_x}), we can get 
{\begin{equation}\label{eq_prob_emb_tgt_comp}
    p(\mathbf{y}_i|\mathbf{e}_i,\mathbf{x}_i)\ge p(\mathbf{y}_i,\mathbf{e}_i|\mathbf{x}_i).
\end{equation}}
{Intuitively, Eq. (\ref{eq_prob_emb_tgt_comp}) implies that} the learning objective in Eq. (\ref{eq_with_prob_emb_tgt}) can be more deterministic and effective than that in Eq. (\ref{eq_no_prob_emb_tgt}). {By considering Eq. (\ref{eq_conditional_e_x}) and Eq. (\ref{eq_prob_emb_tgt_comp}) jointly, we can observe that as the inputs of different problems become more indistinguishable, $p(\mathbf{e}_i|\mathbf{x}_i)$ approaches 0, and the gap between the two learning objectives increases.}

Specifically, due to the uncertain correspondence between \( \mathbf{x} \) and \( \mathbf{e} \), attempting to infer the problem embedding \( \mathbf{e}_i \) directly from \( \mathbf{x}_i \) in Eq. (\ref{eq_no_prob_emb_tgt}) is theoretically flawed and likely to introduce noise. Moreover, assuming the output space of \( \mathbf{y} \) is of size \( Y \), and that of \( \mathbf{e} \) is \( E \) (even though both are infinite in practice), the output space of Eq. (\ref{eq_no_prob_emb_tgt}) becomes \( Y \times E \), which is significantly larger than the output space \( Y \) of Eq. (\ref{eq_with_prob_emb_tgt}). This leads to a sparsity of training samples per output configuration in Eq. (\ref{eq_no_prob_emb_tgt}), making the learning less efficient and more prone to underfitting.

In summary, by explicitly leveraging problem embeddings, Eq. (\ref{eq_with_prob_emb_tgt}) avoids the uncertainty and redundancy inherent in the joint modeling of problem and solution in Eq. (\ref{eq_no_prob_emb_tgt}). This brings a more focused learning objective with stronger generalization capability, making it a theoretically superior choice.

\subsection{Rank Pool for Problem Feature Embedding}\label{sec_rank_pool_method}

The token sequence length of LLM-embedded mathematical formulations varies with the length of the underlying expressions. Although PAD leverages a Transformer-based architecture, such variable-length problem embeddings pose challenges for efficient learning. Hence, a token selection mechanism is required to extract key information into a fixed-length sequence. We observe that raw problem embeddings exhibit three distinct characteristics:
\begin{enumerate}
    \item Structural Sparsity. Mathematical formulations of optimization problems are inherently structured, featuring segmented semantic components, including objective functions, constraints, and optimization variables. However, the corresponding tokens are often distributed sparsely and non-contiguously across the sequence, making it more difficult to retrieve meaningful structure.
    \item Symbol Sensitivity. Small changes in mathematical symbols can lead to entirely different types of problems. Moreover, global operators such as $\mathrm{min}$, $\sum$, or nested formulations require the structural integrity of the expression to be preserved during token selection.
    \item Long-Tail Distribution. Standard low-level tokens (e.g., basic operators or variable names) appear frequently but carry limited semantic value. In contrast, high-information global symbols appear less frequently but are crucial for understanding the problem structure. A robust token selection mechanism must retain and emphasize these rare but informative tokens.
\end{enumerate}

Conventional pooling strategies struggle to handle these properties. For instance, MaxPool \cite{pytorch}, which selects the most active tokens at each dimension, is highly susceptible to noise and often overemphasizes frequent but low-informative symbols (e.g., repeating input variable names), while potentially breaking mathematical structure by discarding crucial operator relationships. Moreover, MaxPool can fragment tokens, resulting in pooled tokens that are combinations of incomplete parts from multiple original tokens. AvgPool \cite{pytorch}, on the other hand, dilutes the semantic strength of key tokens and is insensitive to positional structure, failing to preserve the underlying formulation hierarchy. Although these methods have been widely adopted in other tasks, they demonstrate poor performance when applied to token selection in problem embeddings due to the unique structural and symbolic demands of mathematical optimization representations.

We propose a token selection method based on intra-sequence token similarity scores derived from the problem embedding, referred to as the rank pool. Specifically, given a raw problem embedding sequence of length \( S \), we define a similarity score matrix \( \mathbf{M}_{\mathrm{SI}} \), where each entry quantifies the similarity between a pair of token vectors. We compute pairwise similarities using PyTorch’s built-in \texttt{cosine\_similarity()} function \cite{pytorch}, where values approaching 1 indicate high similarity and those near 0 indicate low similarity. Formally, the similarity matrix is defined as
\begin{equation}\label{eq_matrix_similarity}
    \mathbf{M}_{\mathrm{SI}}[i][j]=\texttt{cosine\_similarity}(\mathrm{token}_i, \mathrm{token}_j),
\end{equation}
where \( \mathrm{token}_i \) is the \( i \)-th token vector in the original sequence.

Next, we compute a similarity sum vector \( \mathbf{V}_{\mathrm{SUM\_SI}} \), where each element \( \mathbf{V}_{\mathrm{SUM\_SI}}[i] \) represents the total similarity score between token \( i \) and all other tokens
\begin{equation}\label{eq_vector_sum_si}
    \mathbf{V}_{\mathrm{SUM\_SI}}[i]=\sum_{j \ne i} \mathbf{M}_{\mathrm{SI}}[i][j].
\end{equation}
{We then sort the tokens} based on their \( \mathbf{V}_{\mathrm{SUM\_SI}} \) values and select the \( m \) tokens with the lowest total similarity scores, corresponding to the most diverse tokens in the sequence. These \( m \) tokens are sequentially concatenated to form the pooled sequence. Here, \( m \) is a tunable hyperparameter chosen to be a small positive integer significantly smaller than the original sequence length.

By selecting the most diverse tokens, the rank pool effectively captures unique and informative symbolic elements, particularly for those who are sensitive to small changes in mathematical expressions. Rank pool preserves the integrity of token vectors without fragmenting or averaging their elements, allowing each complete token to potentially convey more of the original information. Additionally, keeping the selected tokens in their original order helps preserve important structural relationships within the expression. In practice, rank pool can be performed only once when the problem is first encountered, and the resulting pooled sequence can be cached for reuse, thereby avoiding repeated calls to the LLM for encoding and pooling. The experimental details are shown in Section \ref{sec_exp_rank_pool}.

\begin{algorithm}[t]
\small
\caption{Training of encoder-decoder.}
\label{alg_enc_dec}
\KwIn{$N$ training problems and their corresponding datasets $\boldsymbol{\mathcal{D}}=\{\mathcal{D}_1,\ldots,\mathcal{D}_i,\ldots,\mathcal{D}_N\}$, the encoder model $\boldsymbol{\theta}_\mathrm{E}$, the decoder model $\boldsymbol{\theta}_\mathrm{D}$;}
\KwOut{Trained models $\boldsymbol{\theta}_\mathrm{E}$ and $\boldsymbol{\theta}_\mathrm{D}$;}

\Repeat{convergence}{
    \For{$\mathrm{zipped\ batch\ from}\ \boldsymbol{\mathcal{D}}$}{
        \For{$i = 1, \dots, N$}{
          $\mathbf{x},\mathbf{y}$ $\leftarrow$ batch\ data\ from\ $\mathcal{D}_i$\;
          $\mathbf{v}_{\mathrm{joint}}$ $\leftarrow$ cat($\mathbf{x},\mathbf{y}$)\;
          $\tilde{\mathbf{v}}_{\mathrm{joint}}$ $\leftarrow$ $f_{\boldsymbol{\theta}_\mathrm{D}}(f_{\boldsymbol{\theta}_\mathrm{E}}(\mathbf{x}))$\;
          $\mathbf{v}_{\mathrm{val}},\mathbf{v}_{\mathrm{type}}$ $\leftarrow$ split\ $\mathbf{v}_{\mathrm{joint}}$\;
          $\tilde{\mathbf{v}}_{\mathrm{val}},\tilde{\mathbf{v}}_{\mathrm{type}}$ $\leftarrow$ split\ $\tilde{\mathbf{v}}_{\mathrm{joint}}$\;
          $\mathcal{L}_{\mathrm{real}}$ $\leftarrow$ $\| \mathbf{v}_{\mathrm{val}}-\tilde{\mathbf{v}}_{\mathrm{val}} \|^2$\;
          $\mathcal{L}_{\mathrm{class}}$ $\leftarrow$ $\mathrm{CrossEntropyLoss}(\mathbf{v}_{\mathrm{type}},\tilde{\mathbf{v}}_{\mathrm{type}})$\;
          $\mathcal{L}$ $\leftarrow$ $\mathcal{L}_{\mathrm{real}}+\mathcal{L}_{\mathrm{class}}$\;
          Backpropagation\ on\ $\boldsymbol{\theta}_\mathrm{E}$\ and\ $\boldsymbol{\theta}_\mathrm{D}$\ with\ $\mathcal{L}$\;
        }
    }
}
\end{algorithm}

\subsection{Problem-Aware Diffusion}

We design a Transformer-based LDM \cite{StableDiff2022CVPR} as the neural network component of PAD, with the Transformer \cite{attention2017transformer} architecture as its backbone. Specifically, LDM is characterized by its ability to project variable-dimensional inputs from the data space into a fixed-dimensional latent space via an encoder, and then perform diffusion-based noise adding and denoising in this latent space. As illustrated in Fig. \ref{fig_framework}, our element-wise encoder maps each individual element in the input vector \( \mathbf{x} \) and solution vector \( \mathbf{y} \) into a token vector, in a way analogous to how a text encoder tokenizes characters. To account for the distinct physical meanings of different elements in \( \mathbf{x} \) and \( \mathbf{y} \), we assign an integer label to each element representing its physical type (e.g., power, channel gain), allowing the encoder to generate discriminative embeddings. {These physical types are mainly designed to enable the element-wise encoder–decoder to produce parameter and variable embeddings with physical meaning, thereby allowing the latent diffusion model to learn specialized, high-quality solution distributions for data with different physical interpretations. In practice, however, the latent diffusion model is not required to achieve perfect classification of variable types; instead, it focuses on learning the distribution of high-quality variable values, which avoids the excessive difficulty of simultaneously generating both variable values and types.}

Let the encoder and decoder networks be parameterized by \( \boldsymbol{\theta}_\mathrm{E} \) and \( \boldsymbol{\theta}_\mathrm{D} \), respectively. The training objective of the encoder-decoder is given as $f_{\boldsymbol{\theta}_\mathrm{D}}(f_{\boldsymbol{\theta}_\mathrm{E}}(\mathbf{x}))=\mathbf{x}$, where we apply mean squared error (MSE) loss on the element values and cross-entropy loss on the associated physical type labels. The training procedure is detailed in \textbf{Algorithm \ref{alg_enc_dec}}.

\begin{algorithm}[t]
\small
\caption{Training of constraint-aware module.}
\label{alg_cons}
\KwIn{$N$ training problems and their corresponding datasets $\boldsymbol{\mathcal{D}}=\{\mathcal{D}_1,\ldots,\mathcal{D}_i,\ldots,\mathcal{D}_N\}$ and {complete problem embedding after rank pooling $\{\mathcal{E}^\mathrm{p,eq}_1,\ldots,\mathcal{E}^\mathrm{p,eq}_i,\ldots,\mathcal{E}^\mathrm{p,eq}_N\}$,} the constraint-aware module $\boldsymbol{\theta}_\mathrm{C}$, constraint judgment tool function $\mathrm{if\_satisfy\_cons}()$;}
\KwOut{Trained model $\boldsymbol{\theta}_\mathrm{C}$;}

\Repeat{convergence}{
    \For{$\mathrm{zipped\ batch\ from}\ \boldsymbol{\mathcal{D}}$}{
        \For{$i = 1, \dots, N$}{
          $\mathbf{x}$ $\leftarrow$ batch\ data\ from\ $\mathcal{D}_i$\;
          $\mathbf{y}$ $\leftarrow$ randomly\ sampled\ solutions\;
          $\mathbf{e}_i$ $\leftarrow$ randomly\ selected\ from\ $\mathcal{E}^\mathrm{p,eq}_i$\;
          $\tilde{\mathbf{r}}$ $\leftarrow$ $f_{\boldsymbol{\theta}_\mathrm{C}}(\mathbf{e}_i,\mathbf{x},\mathbf{y})$\;
          $\mathbf{r}$ $\leftarrow$ $\mathrm{if\_satisfy\_cons}(\mathbf{e}_i,\mathbf{x},\mathbf{y})$\;
          $\mathcal{L}$ $\leftarrow$ $\mathrm{CrossEntropyLoss}(\mathbf{r},\tilde{\mathbf{r}})$\;
          Backpropagation\ on\ $\boldsymbol{\theta}_\mathrm{C}$\ with\ $\mathcal{L}$\;
        }
    }
}
\end{algorithm}

For the backbone of PAD, under the $i$-th problem, we incorporate the rank pool mechanism introduced in Section \ref{sec_rank_pool_method}. As shown in Fig. \ref{fig_neural}, the encoded sequence of \( \mathbf{x}_i \) is concatenated with the pooled problem embedding sequence obtained via rank pool, forming a condition sequence \( \mathbf{c} \) that guides the denoising process.

During training, the encoded optimal solution sequence \( \mathbf{z}^0 \) is progressively noised according to a diffusion schedule up to time step \( T \), eventually yielding a pure Gaussian noise vector \( \mathbf{z}^T \). The Transformer-based diffusion backbone \( \boldsymbol{\theta}_\mathrm{B} \) takes the noisy vector \( \mathbf{z}^t \) and the condition sequence \( \mathbf{c} \) as input, and predicts the noise to be removed. The model is trained using MSE loss between the predicted and actual noise.

Here, \( \mathbf{c} \) and \( \mathbf{z}^t \) correspond to the source and target sequences for the Transformer, respectively. At inference time, we sample \( \mathbf{z}^T \) directly from Gaussian noise and iteratively apply \( \boldsymbol{\theta}_\mathrm{B} \) to denoise it back to \( \tilde{\mathbf{z}^0} \) without noise, which is then decoded to obtain the final generated solution \( \tilde{\mathbf{y}_i} \).

\begin{algorithm}[t]
\small
\caption{Training of PAD backbone.}
\label{alg_pad}
\KwIn{$N$ training problems and their corresponding datasets $\boldsymbol{\mathcal{D}}=\{\mathcal{D}_1,\ldots,\mathcal{D}_i,\ldots,\mathcal{D}_N\}$ and {complete problem embedding after rank pooling $\{\mathcal{E}^\mathrm{p,eq}_1,\ldots,\mathcal{E}^\mathrm{p,eq}_i,\ldots,\mathcal{E}^\mathrm{p,eq}_N\}$, constraint missing problem embedding after rank pooling $\{\mathcal{E}^\mathrm{p,miss}_1,\ldots,\mathcal{E}^\mathrm{p,miss}_i,\ldots,\mathcal{E}^\mathrm{p,miss}_N\}$,} diffusion step $T$, constraint loss start epoch $e$, the trained models $\boldsymbol{\theta}_\mathrm{C}$, $\boldsymbol{\theta}_\mathrm{E}$ and $\boldsymbol{\theta}_\mathrm{D}$;}
\KwOut{Trained model $\boldsymbol{\theta}_\mathrm{B}$;}

\Repeat{convergence}{
    \For{$\mathrm{zipped\ batch\ from}\ \boldsymbol{\mathcal{D}}$}{
        \For{$i = 1, \dots, N$}{
          $\mathbf{x},\mathbf{y}^*$ $\leftarrow$ batch\ data\ from\ $\mathcal{D}_i$\;
          $\mathbf{c}_{\mathrm{x}},\mathbf{z}_0$ $\leftarrow$ $f_{\boldsymbol{\theta}_\mathrm{E}}(\mathbf{x},\mathbf{y}^*)$\;
          {$\mathbf{e}_i$ $\leftarrow$ randomly\ selected\ from\ $\mathcal{E}^\mathrm{p,eq}_i\bigcup\mathcal{E}^\mathrm{p,miss}_i$\;
          $\mathbf{c}$ $\leftarrow$ $\mathrm{cat}(\mathbf{e}_i,\mathbf{c}_{\mathrm{x}})$\;}
          $\mathbf{t}$ $\leftarrow$ sample time step from $\mathrm{Uniform}(1,T)$\;
          $\boldsymbol{\epsilon}$ $\leftarrow$ sample noise from $\mathcal{N}(0, \mathbf{I})$\;
          $\mathbf{z}_t$ $\leftarrow$ noise adding on $\mathbf{z}_0$ based on $\boldsymbol{\epsilon}$ and $\mathbf{t}$\;
          $\tilde{\boldsymbol{\epsilon}}$ $\leftarrow$ $f_{\boldsymbol{\theta}_\mathrm{B}}(\mathbf{c},\mathbf{t},\mathbf{z}_t)$\;
          $\mathcal{L}_{\mathrm{eps}}$ $\leftarrow$ $\| \boldsymbol{\epsilon}-\tilde{\boldsymbol{\epsilon}} \|^2$\;
          $\mathcal{L}_{\mathrm{cons}}$ $\leftarrow$ $0$\;
          \If{ {in the last $e$ epochs and $\mathbf{e}_i$ from $\mathcal{E}^\mathrm{p,eq}_i$}}{
            $\mathbf{z}_t,\tilde{\boldsymbol{\epsilon}}$ $\leftarrow$ select $\mathbf{z}_t$ and the corresponding $\tilde{\boldsymbol{\epsilon}}$ where $t<\frac{T}{2}$\;
            $\tilde{\mathbf{z}_0}$ $\leftarrow$ denoising on $\mathbf{z}_t$ using $\tilde{\boldsymbol{\epsilon}}$\;
            $\mathcal{L}_{\mathrm{cons}}$ $\leftarrow$ $f_{\boldsymbol{\theta}_\mathrm{C}}(\mathbf{e}_i,\mathbf{x},f_{\boldsymbol{\theta}_\mathrm{D}}(\tilde{\mathbf{z}_0}))$\;
          }
          $\mathcal{L}$ $\leftarrow$ $\mathcal{L}_{\mathrm{eps}}+\mathcal{L}_{\mathrm{cons}}$\;
          Backpropagation\ on\ $\boldsymbol{\theta}_\mathrm{B}$\ with\ $\mathcal{L}$\;
        }
    }
}
\end{algorithm}

\subsection{Constraint-Aware Module}

{To improve the likelihood of generating feasible solutions, we design an independent constraint-aware module based on a TransformerEncoder \cite{agi2023incentive} and a classification head, which is used only during training.} This module functions as a binary classifier that takes the parameter \( \mathbf{x} \), the solution \( \mathbf{y} \), and the pooled problem embedding as input, and predicts whether \( \mathbf{y} \) violates the hard constraints of the corresponding problem (outputting $1$ if it does, and $0$ otherwise). By approximating non-differentiable constraint functions through a differentiable neural network, this module introduces constraint violation penalties during the solution generation process, guiding the model toward producing valid solutions. This also increases the diversity of feasible solutions discovered during parallel sampling. Augmenting the GDMs' noise prediction loss with additional preference-based objectives has proven effective in the AI community \cite{CLIP}. The training process of the constraint-aware module is detailed in \textbf{Algorithm \ref{alg_cons}}.

In implementation, we prepend a classification (CLS) token \cite{BERT} to the input sequence and use its output representation for downstream classification. To mitigate the potential gradient interference between the noise prediction loss \( \mathcal{L}_{\mathrm{eps}} \) and the constraint violation loss \( \mathcal{L}_{\mathrm{cons}} \), we apply \( \mathcal{L}_{\mathrm{cons}} \) only during the final phase of training. By introducing \( \mathcal{L}_{\mathrm{cons}} \) after \( \mathcal{L}_{\mathrm{eps}} \) has largely converged, we enable more efficient optimization of \( \mathcal{L}_{\mathrm{cons}} \) while avoiding the convergence instability that may arise from joint training from the beginning. Moreover, we compute \( \mathcal{L}_{\mathrm{cons}} \) only for the final solutions derived from the denoising steps in the latter half of the diffusion process. This strategy ensures that the incorporation of constraints does not excessively disturb the denoising trajectory.

{To enhance the model's sensitivity to constraints and robustness to the diversity of problem embeddings, we design a constraint-gated loss training mechanism inspired by the idea of classifier-free guidance (CFG) \cite{ho2022classifier}. This mechanism prevents the constraint signal from being treated merely as background information of input parameters and thus overlooked by the model. Specifically, for each problem ($i$), we construct two sets of embeddings during training, denoted as $\mathcal{E}^{\mathrm{p,eq}}_i$ and $\mathcal{E}^{\mathrm{p,miss}}_i$. All embeddings in $\mathcal{E}^{\mathrm{p,eq}}_i$ are equivalent to the original problem, differing only in the ordering of terms in the objective and constraint functions or in variable signs. In contrast, the embeddings in $\mathcal{E}^{\mathrm{p,miss}}_i$ not only involve such permutations and sign variations, but also include incomplete constraint specifications, where parts or even all of the constraints may be missing. Both $\mathcal{E}^{\mathrm{p,eq}}_i$ and $\mathcal{E}^{\mathrm{p,miss}}_i$ contain 3–5 different embeddings each, which are used exclusively during training. At each iteration, the model randomly samples an embedding from their union, and the constraint-aware module performs forward computation of the constraint loss only when the input comes from $\mathcal{E}^{\mathrm{p,eq}}_i$. This also reduces the computational overhead of automatic differentiation and backpropagation through the constraint-aware module. It is worth noting that our constraint-gated loss differs fundamentally from CFG. The constraint-gated loss is determined solely by the content of the diffusion model's input condition rather than by the presence or absence of a problem embedding. Our design is primarily intended as an exploration of a novel training strategy to strengthen the model's awareness of constraints, rather than to reproduce the expected effects of CFG's dual noise-prediction and weighted-sum mechanism. During the inference phase, embeddings are introduced only from $\mathcal{E}^{\mathrm{p,eq}}_i$.}

Once the encoder-decoder and constraint-aware module are trained, the overall training process of the PAD backbone is described in \textbf{Algorithm \ref{alg_pad}}, where all parameters outside the backbone are kept frozen.

\section{Experiments and Discussions}

\subsection{Experimental Settings}

\begin{table}[t]
    \centering
    \captionsetup{font={small}}
    \caption{{Default settings for PAD's main hyperparameters.}}
    \begin{tblr}{
      colspec={X[1.2,l] X[0.5,c]}, 
      cells = {c},
      hlines,
      vlines,
      rowsep = {2pt},
    }
    \textbf{Model Parameter} & \textbf{Value} \\ 
    \texttt{encode\_dims}   & [4, 32, 128] \\
    \texttt{hidden\_dim}   & 128 \\
    \texttt{decode\_dims}  & [128, 32] \\
    \texttt{type\_num}  & 19 \\
    \texttt{encoder\_decoder\_train\_epochs}  & 20 \\
    \texttt{encoder\_decoder\_init\_lr}  & $1.5\times10^{-4}$ \\
    \texttt{transformer\_encoder\_num}  & 6 \\
    \texttt{transformer\_decoder\_num} & 6 \\
    \texttt{attention\_header\_num} & 8 \\
    \texttt{transformer\_dropout} & 0.1 \\
    \texttt{constraint\_module\_train\_epochs} & 30 \\
    \texttt{constraint\_module\_init\_lr}  & $1.5\times10^{-4}$ \\
    \texttt{pool\_type} & rank pool \\
    \texttt{pool\_length} ($m$) & 20 \\
    \texttt{diffusion\_type} & Gaussian \\
    \texttt{diffusion\_steps} & 50 \\
    \texttt{inference\_diffusion\_steps} & 5 \\
    \texttt{parallel\_sampling\_num} ($p_{\rm num}$) & 2 \\
    \texttt{constrain\_loss\_start\_ratio} & 0.5 \\
    \texttt{use\_gated\_constraint\_loss} & True \\
    \texttt{diffusion\_training\_epochs} & 70 \\
    \texttt{diffusion\_init\_lr}  & $2\times10^{-4}$ \\
    \end{tblr}
    \label{tab_model_configs}
    \vspace{-0.45cm}
\end{table}

\subsubsection{Datasets}\label{sec_datasets}
{We label the ten problems in Section \ref{sec_selected_problems} sequentially as [P1, ..., P10]. For each problem, we set three output variable dimensions and generate optimal samples using the Gurobi solver \cite{gurobi}. Except for P7, where the three dimensions are set to 3, 4, and 5 (corresponding to total output dimensions of 6, 8, and 10) due to the excessive runtime of the solver on higher-dimensional inputs, all other problems use dimensions of 5, 6, and 7 (corresponding to total output dimensions of 10, 12, and 14 in the mixed-integer case). The values of input parameters are configured with reference to realistic settings from related literature, as detailed in our open-source repository.}

{The $\mathcal{E}^\mathrm{p,eq}_i$ and $\mathcal{E}^\mathrm{p,miss}_i$ are obtained by first embedding each LaTeX-format expression with MathBERT \cite{mathbert} to produce raw problem embeddings, and then applying pooling operations.} The resulting embeddings typically range from 100 to 200 tokens in length, with each token having a dimensionality of 768. Although the token sequence lengths of DeepSeekMath \cite{deepseek2024math} are comparable to those of MathBERT, the dimensionality per token reaches 4096 due to differences in its tokenizer or vocabulary. This leads to a prohibitively high computational cost in the token compression layer after pooling, and thus DeepSeekMath is omitted from subsequent experiments. Unless otherwise specified, we adopt our proposed rank pool method with $m=20$ for pooling the problem embeddings in the following experiments.

{To analyze the cross-problem generalization ability of the model, we follow the clustering in Section \ref{sec_selected_problems} and select one problem from each cluster (namely P1, P4, and P7) together with the independent P10 to form the complete training dataset. Each problem contributes 60,000 samples per dimension. Since the difficulty variations within each cluster are not significant, choosing P1, P4, and P7 as representatives is an empirical decision, but it is sufficient to control for variables and support the intended experiments. For evaluation, we use 2,000 non-overlapping test samples per dimension for every problem.}

All datasets were generated on a machine equipped with an Intel(R) i7-13700F processor and 16 GB of RAM. The problem embedding generation, model training, and inference were all performed on an RTX 4090 24 GB.

\subsubsection{Model Settings}
{Table \ref{tab_model_configs} presents the main hyperparameter settings of PAD. Unless otherwise specified, all experimental figures and tables are based on these configurations. The settings of diffusion-related parameters such as \texttt{diffusion\_steps} and \texttt{inference\_diffusion\_steps} in Table \ref{tab_model_configs} are derived from existing engineering analyses on the impact of hyperparameters in diffusion-based solution generation \cite{difusco,liang2025iotj}. For a more detailed understanding, this subsection provides additional descriptions of the encoder–decoder, the constraint-aware module, and the latent diffusion setup.}

For the encoder–decoder, a series of multilayer perceptrons are used to map input parameter values into a high-dimensional latent space (dimension denoted as \texttt{hidden\_dim}) or decode from it. The entries \texttt{encode\_dims} and \texttt{decode\_dims} in Table \ref{tab_model_configs} describe the progressive encoding and decoding dimensions, from 1D parameter values to the \texttt{hidden\_dim} latent representation. Integer-type input parameters are handled via an \texttt{nn.Embedding} layer. Across the ten problems, there are 19 distinct parameter and variable types with different physical meanings, as detailed in our open-source repository. The parameter values are predicted using a Sigmoid activation, while parameter types are predicted using a Softmax activation. {Due to space limitations and the auxiliary role of the encoder–decoder, we did not dedicate a separate section to its performance results. Here, we only note that its element-value MSE loss during testing is approximately $10^{-7}$, and the prediction accuracy of physical types is close to 1.0.}

For the constraint-aware module, we employ a TransformerEncoder backbone. Since uniformly sampled solutions for certain problems may exhibit severe imbalance between valid and invalid samples (e.g., over 80\% of the training samples violate constraints), which harms generalization, we apply oversampling during training to balance the distribution. {Moreover, after each training epoch, we update the cross-entropy loss weights based on the model’s performance in the previous epoch: the class of 0-1 with poorer performance is assigned a higher weight (up to 1.5), while the better-performing class is assigned a lower weight (down to 1).}

For the PAD backbone, we employ a full Transformer architecture. Diffusion noise is scheduled using a cosine scheduler, and denoising diffusion implicit models (DDIM) \cite{song2022ddim} acceleration is applied during sampling. In practice, many existing solution generation methods require tool-integrated reasoning or decoding strategies to further interpret the decoded solutions \cite{AI4MathFrontier,difusco,crossProblemCO2025}. Our approach introduces a simple outside decoding step based on the absolute magnitude of each generated solution element prior to de-normalization, without relying on any external reasoning tools. Constraint losses are introduced only during the final epochs of training; the hyperparameter \texttt{constrain\_loss\_start\_ratio} in Table \ref{tab_model_configs} specifies the proportion of the training process during which the constraint loss is applied. For example, (\texttt{constrain\_loss\_start\_ratio}=0.5) indicates that the constraint loss is incorporated throughout the latter half of training. This setting is empirical: introducing the loss too early may cause gradient competition that hinders convergence of the diffusion loss, while introducing it too late may result in undertraining.

{The batch sizes for the encoder–decoder, constraint-aware module, and PAD’s latent diffusion backbone are 256, 128, and 352, respectively. These settings are primarily determined by hardware limitations (notably GPU memory) and aim to maximize training efficiency, so they are not included in Table \ref{tab_model_configs} due to their limited generalizability. The constraint-aware module uses the smallest batch size because oversampling doubles the number of samples per problem dimension relative to the original dataset, thereby consuming more GPU memory. For learning rates, we employ a cosine annealing schedule across all models. In addition, to mitigate gradient conflicts between different problems, we aggregate losses across all training problems in a single batch and perform backpropagation jointly, rather than using gradient descent on single-problem batches.}

{For the parallel sampling ($p_{\rm num}$) setup during inference, the value of 2 in Table \ref{tab_model_configs} is chosen to amplify the impact of different experimental variables on solution quality. If the parallel sampling count were increased, the performance differences in the figures would become less pronounced, making them harder to observe and analyze.}

\subsubsection{Metrics}
To evaluate the quality of solutions generated by PAD, we define four metrics:
\begin{itemize}
    \item GT\_GAP: This metric directly measures solution quality by computing the absolute difference between 1 and the ratio of the generated solution’s objective value to the ground-truth optimal value, i.e., $\mathrm{GT\_GAP}=|\frac{f(\mathbf{x},\tilde{\mathbf{y}})}{f(\mathbf{x},\mathbf{y})}-1|$. {The final score is the average GT\_GAP across all samples in the test set of one problem.}
    \item {GLOB\_GT\_GAP: For more concise reporting, this simplified version of GT\_GAP directly refers to the average GT\_GAP across all test problems.}
    \item CONS\_ACC: This evaluates constraint satisfaction across all generated solutions from parallel sampling. It is defined as the overall constraint satisfaction rate over all generated samples.
    \item CONS\_IF: This metric evaluates whether at least one valid solution exists among the parallel samples for each input. A value of 1 is assigned if any valid solution exists, and 0 otherwise. {The final score is the average over all test inputs of one problem.}
\end{itemize}

\begin{figure}[t]
\centering
\captionsetup{font={small}}
\centerline{\includegraphics[width=3.65in]{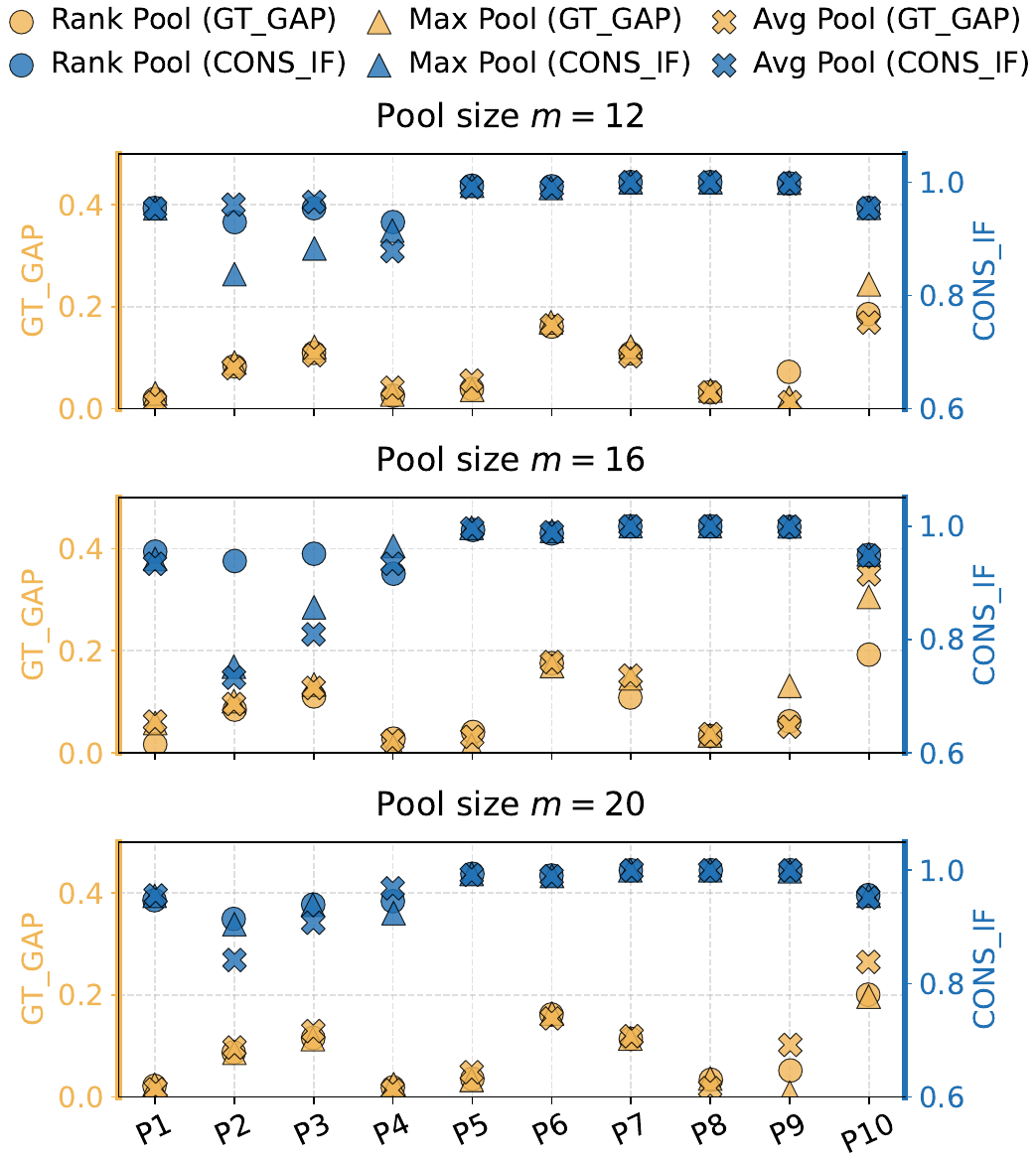}}
\caption{{Comparison of different pooling methods on GT\_GAP $\downarrow$ and CONS\_IF $\uparrow$. The three subplots correspond to the three pooling lengths. Yellow markers represent GT\_GAP (measured on the left y-axis), while blue markers represent CONS\_IF (measured on the right y-axis).}}
\label{fig_pool_method_exp}
\vspace{-0.38cm}
\end{figure}

To further analyze the performance of the constraint-aware module as a binary classifier, we use three standard metrics to assess its classification accuracy: global classification accuracy (ACC), true positive rate (TPR), recall for invalid solutions, and true negative rate (TNR), recall for valid solutions.

\begin{table}[thb]
\centering
\captionsetup{font={small}}
\caption{GLOB\_GT\_GAP $\downarrow$ performance across different pooling methods and pooling lengths ($m$). Here, “Total” denotes the test results aggregated over all problems, “Train” refers to the test results on training problems only, and “Test” indicates the results on problems outside the training set.}
\begin{tabular}{l@{\hskip 0.26in}c@{\hskip 0.15in}c@{\hskip 0.15in}c@{\hskip 0.15in}}\label{tab_total_gt_gap}
 & Rank Pool & Max Pool & Avg Pool \\
\midrule
Total, $m=12$ & $9.58\times10^{-2}$ & $8.00\times10^{-2}$ & $9.58\times10^{-2}$ \\
Total, $m=14$ & $8.00\times10^{-2}$ & $11.08\times10^{-2}$ & $11.06\times10^{-2}$ \\
Total, $m=20$ & $\mathbf{7.64\times10^{-2}}$ & $8.98\times10^{-2}$ & $7.73\times10^{-2}$ \\
\hline
Train, $m=12$ & $8.85\times10^{-2}$ & $9.00\times10^{-2}$ & $10.29\times10^{-2}$ \\
Train, $m=16$ & $8.62\times10^{-2}$ & $13.06\times10^{-2}$ & $14.65\times10^{-2}$ \\
Train, $m=20$ & $8.45\times10^{-2}$ & $10.58\times10^{-2}$ & $\mathbf{8.13\times10^{-2}}$ \\
\hline
Test, $m=12$ & $10.07\times10^{-2}$ & $7.33\times10^{-2}$ & $9.11\times10^{-2}$ \\
Test, $m=16$ & $7.59\times10^{-2}$ & $9.75\times10^{-2}$ & $8.66\times10^{-2}$ \\
Test, $m=20$ & $\mathbf{7.11\times10^{-2}}$ & $7.92\times10^{-2}$ & $7.46\times10^{-2}$ \\
\bottomrule
\end{tabular}
\vspace{-0.45cm}
\end{table}

\subsection{Analysis of Rank Pool}\label{sec_exp_rank_pool}
\subsubsection{Specific Settings}
{To empirically validate the effectiveness of the proposed rank pool method for extracting fixed-length problem embeddings, we conduct experiments using three different pooling methods—Rank Pool, MaxPool, and AvgPool—to directly compare their impact on solution quality. In addition, we design three groups of experiments with different pooling lengths ($m=12,16,20$) to analyze how it affects solution quality. Except for variations in the pooling method and pooling length, all models are constructed according to the settings in Table \ref{tab_model_configs}, with the constraint-aware module introduced under the same configuration.}

\begin{figure}[t]
\centering
\captionsetup{font={small}}
\centerline{\includegraphics[width=3.56in]{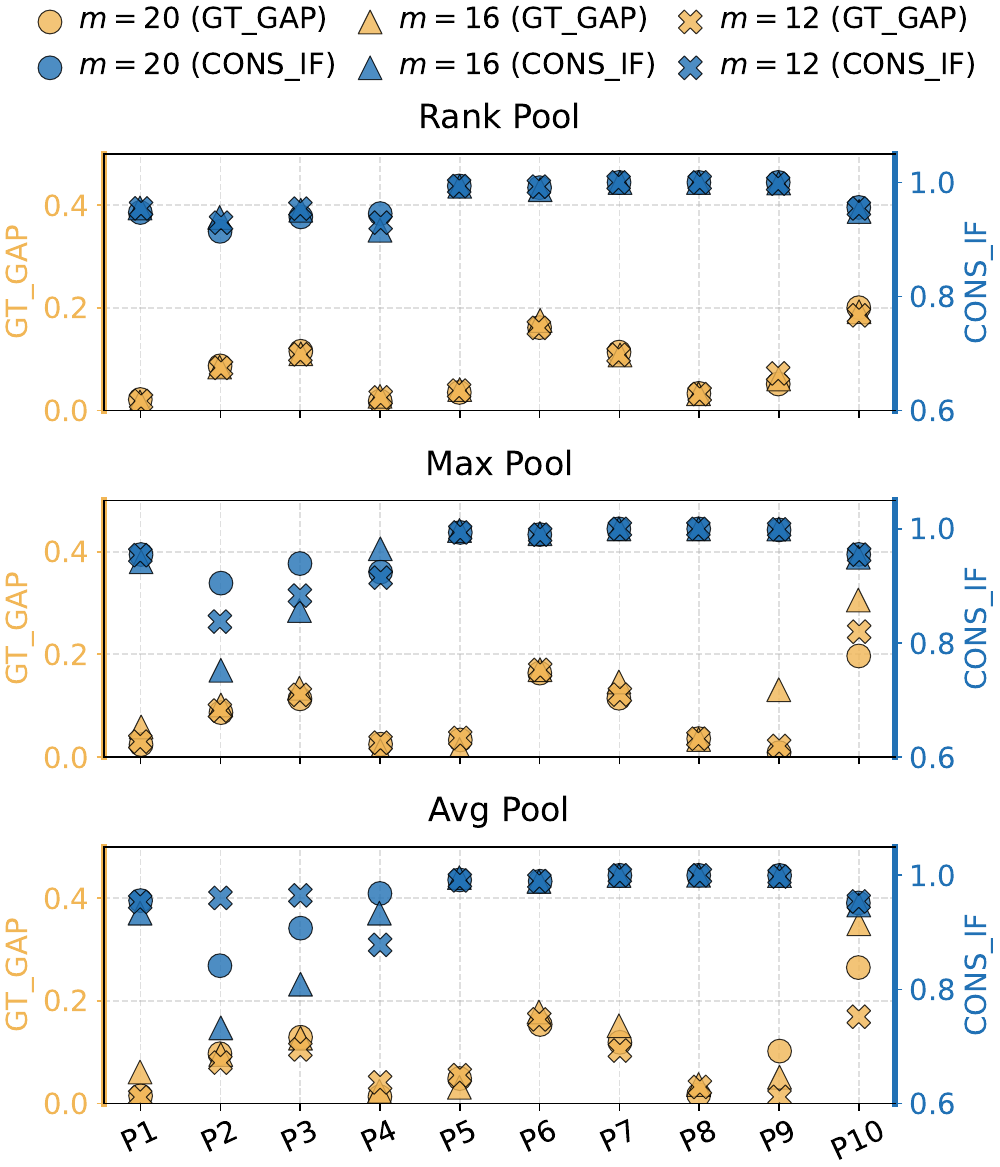}}
\caption{{Comparison of different pooling lengths on GT\_GAP $\downarrow$ and CONS\_IF $\uparrow$. The three subplots correspond to the three pooling methods. Yellow markers represent GT\_GAP (measured on the left y-axis), while blue markers represent CONS\_IF (measured on the right y-axis).}}
\label{fig_pool_size_exp}
\vspace{-0.48cm}
\end{figure}

\begin{figure*}[t]
\centering
\captionsetup{font={small}}
\centerline{\includegraphics[width=5.65in]{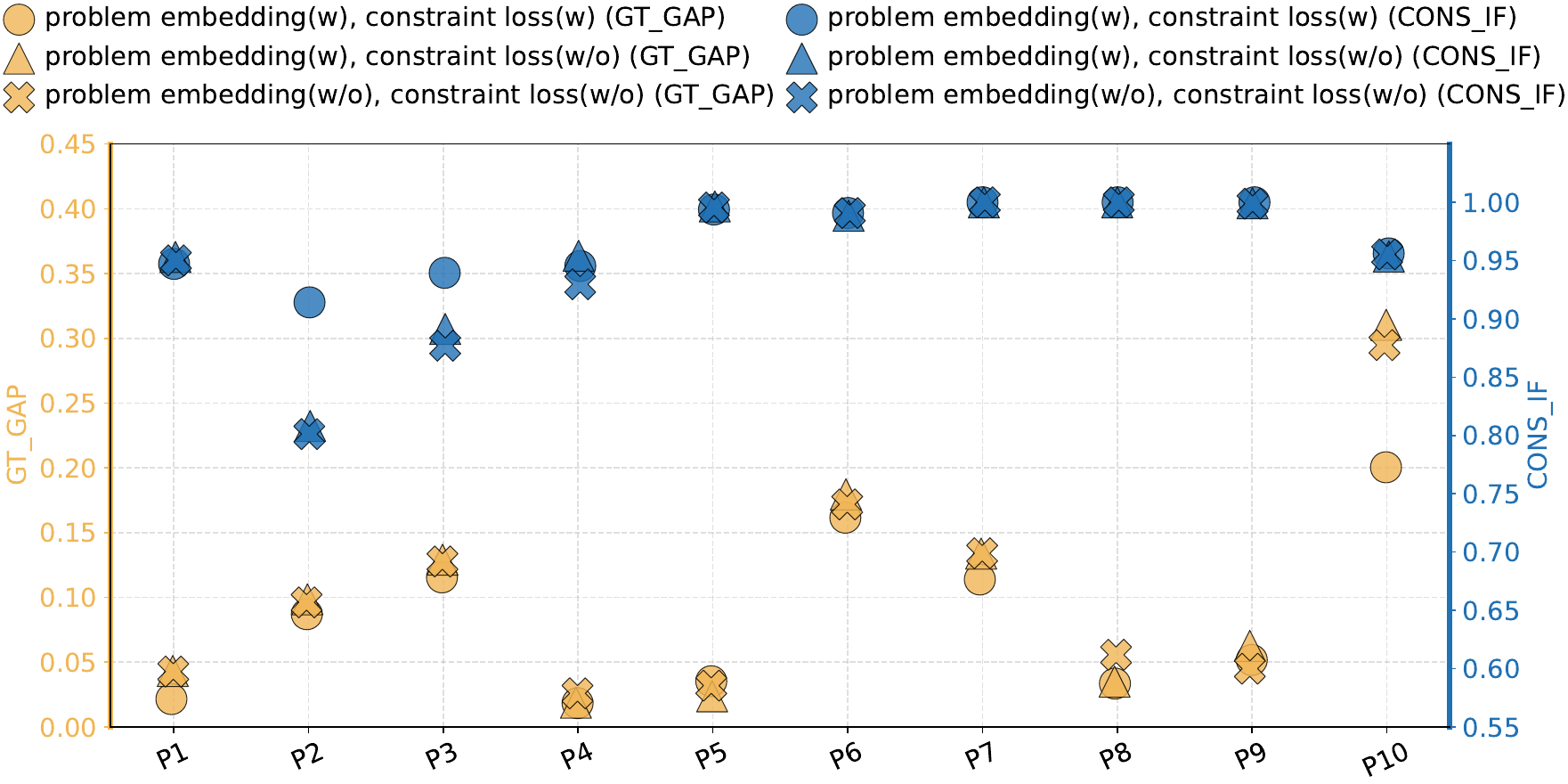}}
\caption{Comparison of with and without problem embeddings and constraint loss on GT\_GAP $\downarrow$ and CONS\_IF $\uparrow$.}
\label{fig_with2without_prob_cons}
\vspace{-0.48cm}
\end{figure*}

\subsubsection{Results} 
{The comparative results for pooling methods and pooling lengths are presented in Table \ref{tab_total_gt_gap}, Fig. \ref{fig_pool_method_exp} and \ref{fig_pool_size_exp}.}

{Table \ref{tab_total_gt_gap} highlights the clearest comparison between pooling methods and lengths. Regarding pooling length, both rank pool and AvgPool exhibit a trend where larger $m$ consistently improves performance. In contrast, MaxPool behaves unusually. The MaxPool performs worst at $m=16$ but best at $m=12$, which may be attributed to the embedding characteristics of the selected problems, where fragmenting tokens through pooling yields some distinguishable representations at smaller $m$. In terms of pooling methods, rank pool achieves the best performance on global problems (“Total”) and unseen problems (“Test”) as $m$ increases, but is slightly outperformed by AvgPool on training problems (“Train”). This may be because once $m$ becomes sufficiently large, information is already well preserved, reducing the negative learning effect of AvgPool’s smoothing operation and allowing it to achieve a better balance.}

\subsection{Effectiveness of Problem Embedding}\label{sec_exp_prob_emb}
\subsubsection{Specific Settings}
{To illustrate the effectiveness of incorporating problem feature embedding, we conduct experiments under the default training setup of Table \ref{tab_model_configs}, exploring the impact of including or excluding problem feature embedding on performance.}

{Fig. \ref{fig_pool_method_exp} and Fig. \ref{fig_pool_size_exp} present dual-metric comparisons of GT\_GAP and CONS\_IF across different problems with respect to pooling methods and pooling lengths, respectively. However, due to variations in problem difficulty and the randomness introduced when $p_{\rm num}=2$, the overall advantages are not immediately obvious, and these figures serve better as references for observing the overall performance range. In Fig. \ref{fig_pool_method_exp}, the Rank pool is seen to achieve the best results on both GT\_GAP and CONS\_IF for most problems. Fig. \ref{fig_pool_size_exp} shows a trend largely consistent with Table \ref{tab_model_configs}, where only the rank pool exhibits a tendency for performance to improve as $m$ increases.}

\subsubsection{Summary} {These findings highlight the effectiveness of pooling original problem feature embeddings into a fixed length under limited model size and computational resources, providing empirical guidance for the choice of pooling methods and pooling lengths.}

Overall, the rank pool method offers a lightweight yet effective strategy for compressing long, symbolic representations into fixed-length embeddings. {Its advantages lie not only in capturing critical token-level information but also in offering better support for the model’s final solution generation, particularly demonstrating stronger performance on unseen problems outside the training set.}

\subsubsection{Results}
{As shown in Fig. \ref{fig_with2without_prob_cons}, on GT\_GAP, the model with problem embedding (yellow triangles) outperforms the one without problem embedding (yellow crosses) on most problems, particularly demonstrating better generalization on out-of-training problems (P5 and P8). On P6, P9, and P10, however, the results without problem embedding are slightly better. We attribute this to the limited model capacity, which amplifies gradient competition across different problems, thereby increasing the learning difficulty when problem embeddings are included and leading to weaker performance in certain cases. In conjunction with the findings of Section \ref{sec_theory_basics}, we conjecture that P6, P9, and P10 may inherently allow the problem structure to be inferred directly from the input parameters. As a result, excluding problem embedding reduces training complexity while having only a minor impact on problem identification. For CONS\_IF, neither the model with problem embedding (blue triangles) nor the one without it (blue crosses) incorporates constraint loss. Nonetheless, the explicit mathematical structure and problem-discriminative information introduced by problem embeddings improve the feasibility rate of generated solutions on some problems.}

\subsubsection{Summary}
Overall, the results in Fig. \ref{fig_with2without_prob_cons} demonstrate that incorporating problem feature embedding effectively enhances the model's ability to generalize across different problems. This improvement is evident in both the quality and feasibility of the generated solutions, highlighting the potential of problem-aware learning.

\subsection{Performance of Constraint-Aware Module}\label{sec_exp_cons_module}

\begin{figure}[t]
\centerline{\includegraphics[width=3.35in]{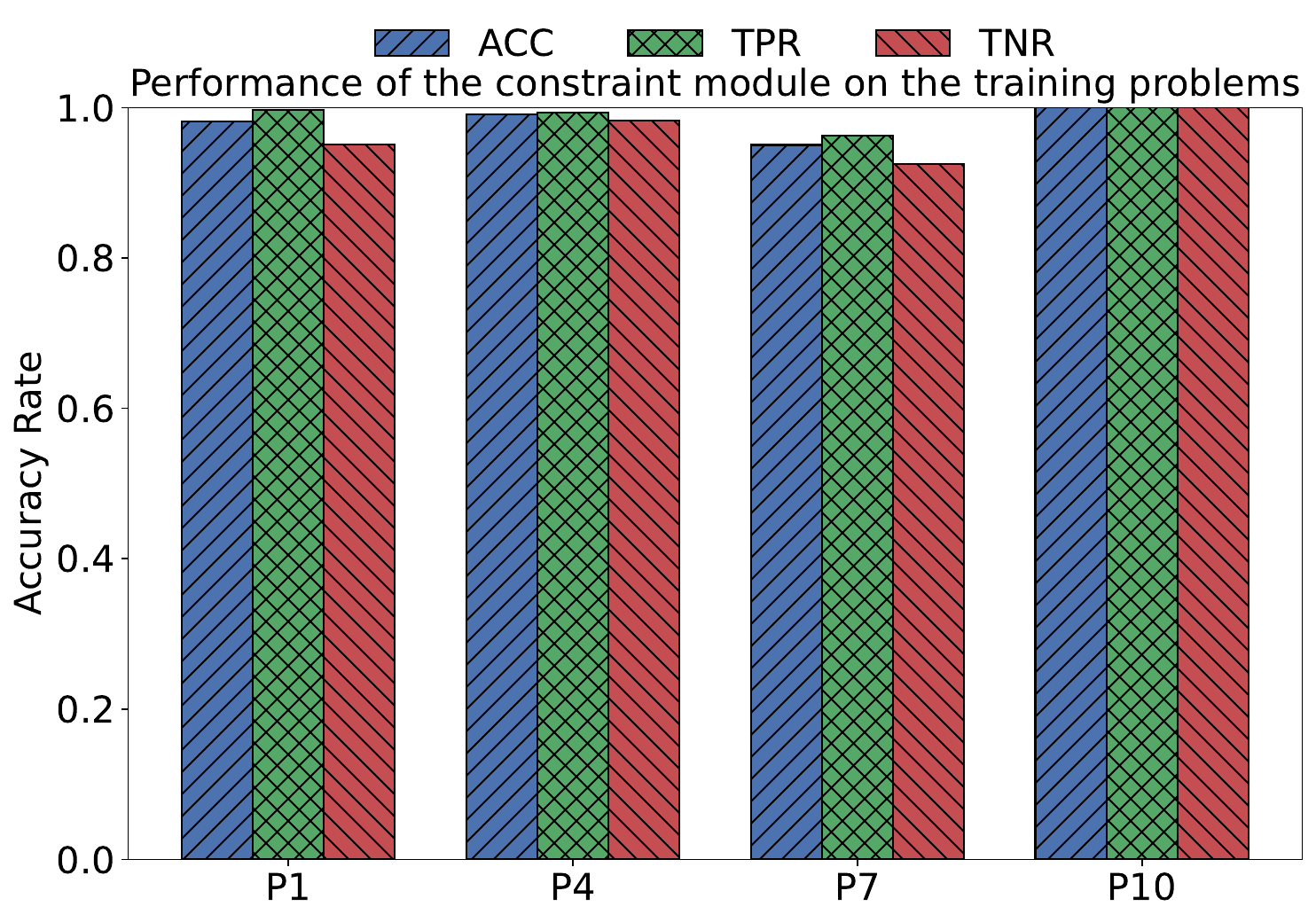}}
\captionsetup{font={small}}
\caption{{Performance of the constraint-aware module in terms of ACC $\uparrow$, TPR $\uparrow$, and TNR $\uparrow$ on training problems.}}
\label{fig_cons_module}
\vspace{-0.38cm}
\end{figure}

\begin{figure}[t]
\centering
\captionsetup{font={small}}
\centerline{\includegraphics[width=3.6in]{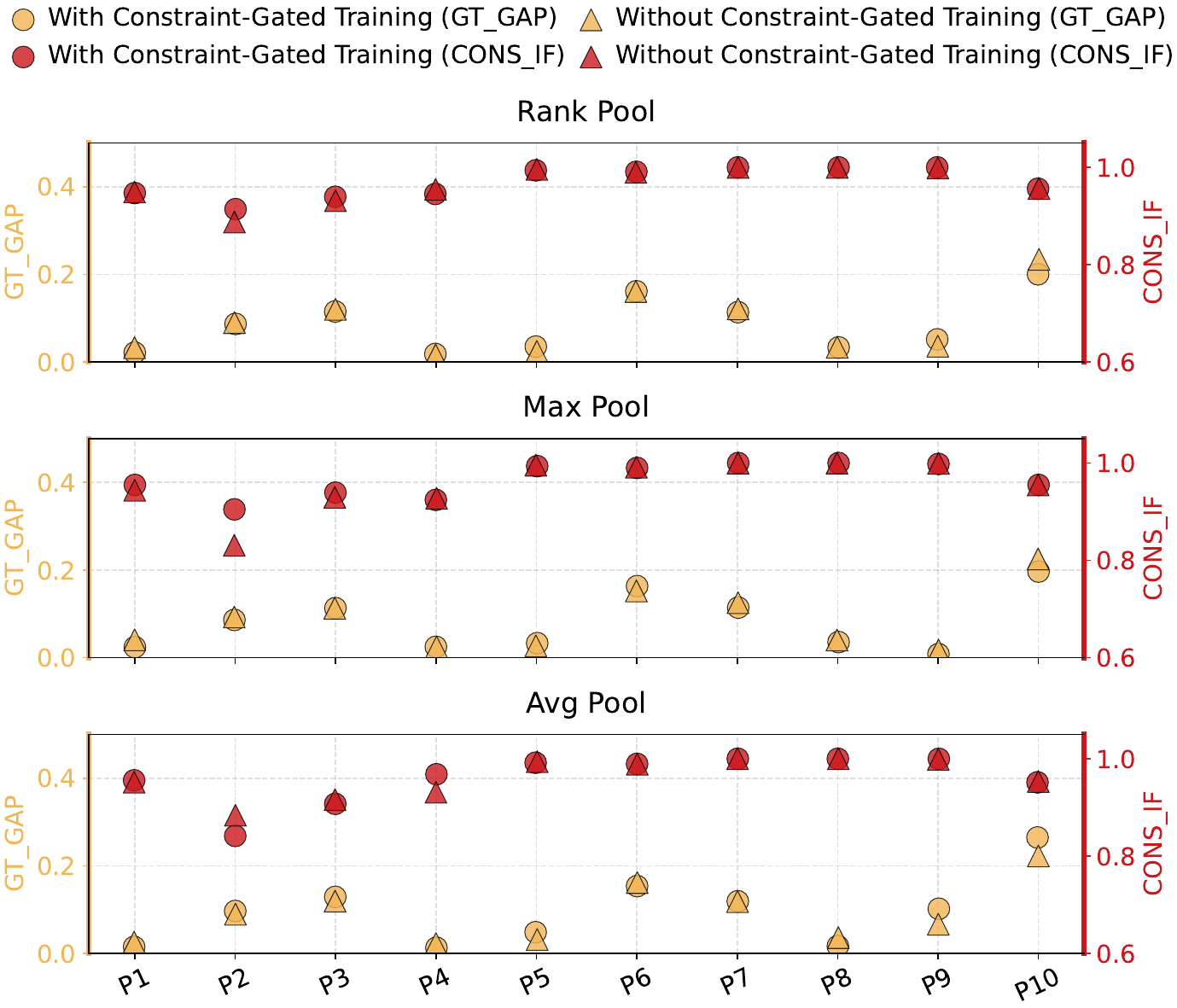}}
\caption{{Comparison of with and without the constraint-gated training mechanism on GT\_GAP $\downarrow$ and CONS\_IF $\uparrow$.}}
\label{fig_constrain_gated_exp}
\vspace{-0.38cm}
\end{figure}

\subsubsection{Specific Settings}
{To demonstrate the effectiveness of the constraint-aware module, we compare the PAD model's performance with and without introducing the constraint loss. Furthermore, we conduct ablation studies on the constraint-gated loss mechanism across all three pooling strategies to assess its effectiveness. Additionally, we illustrate the constraint satisfaction classification performance of the shared frozen constraint-aware module using three metrics: ACC, TPR, and TNR.}

\subsubsection{Results}
{As shown in Fig. \ref{fig_with2without_prob_cons}, when comparing models trained with constraint loss (yellow circles) and without constraint loss (yellow triangles), we observe a clear GT\_GAP improvement across almost all problems, with only a few cases showing no gain, likely due to limited model capacity. This improvement primarily stems from the significantly higher proportion of feasible solutions generated after introducing constraint loss into the training process. This can also be observed from the comparison of CONS\_IF under constraint loss (blue circles) and without it (blue triangles). The increased feasibility ratio directly enhances the efficiency of parallel sampling in exploring high-quality solution distributions, thereby yielding a stable improvement in overall solution quality.}

{As shown in Fig. \ref{fig_cons_module}, our constraint-aware module is used only during training, and thus we report its discrimination accuracy on the four training problems. The module achieves accuracy above 0.9 across all three metrics—ACC, TPR, and TNR. As a differentiable surrogate for the original non-differentiable constraint functions, such accuracy is already sufficient. High TPR and TNR help minimize the risk of penalizing feasible solutions or overlooking infeasible ones, thereby preventing the diffusion model from learning incorrectly.}

{As shown in Fig. \ref{fig_constrain_gated_exp}, introducing the constraint-gated loss mechanism (yellow circles) yields only marginal GT\_GAP improvement compared with not using it (yellow triangles), and for rank pool and AvgPool it even leads to opposite effects on certain problems (P9, P10). We suspect this is because half of the training resources originally allocated to full problem embedding and constraint loss are diverted to incomplete embeddings with constraint loss disabled, resulting in insufficient training under full constraints. However, the mechanism shows significant improvements on CONS\_IF in certain problems, indicating its effectiveness.}

\begin{figure}[t]
\captionsetup{font={small}}
\centerline{\includegraphics[width=3.53in]{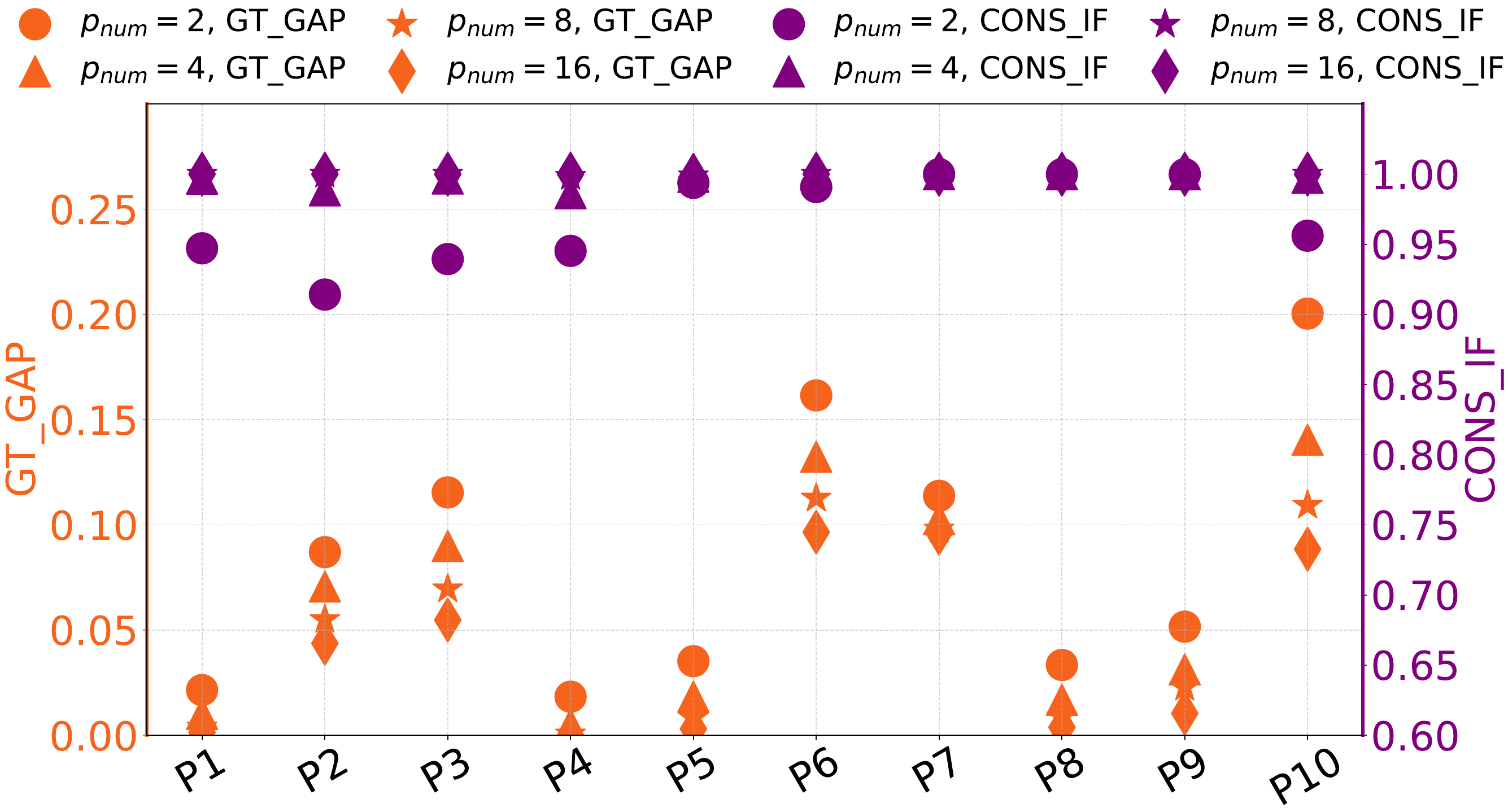}}
\caption{{Effect of parallel sampling $p_{num}$ on GT\_GAP $\downarrow$ and CONS\_IF $\uparrow$.}}
\label{fig_parallel_sampling_exp}
\vspace{-0.38cm}
\end{figure}

\subsubsection{Summary}
{In summary, the results presented in Fig. \ref{fig_with2without_prob_cons}, Fig. \ref{fig_cons_module} and \ref{fig_constrain_gated_exp} demonstrate that incorporating constraint loss effectively improves the feasibility of the solutions generated by the PAD model, while also contributing to better overall solution quality. Furthermore, although the constraint-gated loss does not yield significant improvement on GT\_GAP, its positive effect on CONS\_IF remains empirically valuable. These findings validate the effectiveness of the constraint-aware module and provide an explorative direction for enhancing the model’s sensitivity to constraint functions within problem embeddings.}

\begin{table}[thb]
    \setlength{\tabcolsep}{0.4pt}
    \centering
    \captionsetup{font={small}}
    \caption{\textcolor{black}{Efficiency comparison between PAD, the LLM-based solver OPRO \cite{deepmind2024LLMasOptimizer}, and Gurobi \cite{gurobi}. The “Single-sample Inference Time” represents the lower and upper bounds of the latency observed for each method across all problems and input dimensions. ``-" indicates metrics that cannot be precisely obtained or are not applicable to the corresponding method.}}
    \begin{tabularx}{\linewidth}{|>{\raggedright\arraybackslash}m{0.20\linewidth}
                                   |>{\centering\arraybackslash}m{0.26\linewidth}
                                   |>{\centering\arraybackslash}m{0.26\linewidth}
                                   |>{\centering\arraybackslash}m{0.26\linewidth}|}
    \hline
    \textbf{Property} & \textbf{PAD ($p_{num}=2$)} & \textbf{OPRO (GPT-5)} & \textbf{Gurobi} \\
    \hline
    GT\_GAP $\downarrow$                     & $0.0083\sim0.2229$ & $0.000\sim0.8276$ & $0.0000$ \\
    \hline
    CONS\_IF $\uparrow$                    & $0.85\sim1.00$ & $0.50\sim1.00$ & $1.00$ \\
    \hline
    Training /Building Time               & 2.78h & -- & Depends on manual coding \\
    \hline
    Single-sample Inference Time& 7.19ms$\sim$164.47ms & 1.7s$\sim$5.43s & 1.09ms$\sim$917.43ms \\
    \hline
    Model Parameter Quantity                  & 2.04 Million & 52 Trillion & -- \\
    \hline
    Runtime Processor Utilization    & 48\% (GPU) & -- & $10\%\sim95\%$ (CPU) \\
    \hline
    Runtime Memory Usage    & 1.55GB (GPU) & -- & 0.3GB (CPU) \\
    \hline
    Hardware           & 1$\times$ RTX 4090 & -- & Intel(R) i7-13700F 16 GB RAM \\
    \hline
    \end{tabularx}
    \label{tab_method_efficency}
    \vspace{-0.48cm}
\end{table}

\begin{table*}[t]
\centering
\captionsetup{font={small}}
\caption{\textcolor{black}{The impact of removing one problem from the four training problems on the PAD model’s GLOB\_GT\_GAP $\downarrow$ performance. Each column corresponds to the GLOB\_GT\_GAP performance on a different cluster or an individual problem. For models trained with one problem removed, a (+) indicates that the score is better than that of the full training model, while a (–) indicates it is worse.}}
\begin{tabular}{l@{\hskip 0.26in}c@{\hskip 0.2in}c@{\hskip 0.2in}c@{\hskip 0.2in}c@{\hskip 0.2in}}\label{tab_training_problem_ablation}
 & P1, P2, P3 & P4, P5, P6 & P7, P8, P9 & P10 \\
\midrule
Trained on P1, P4, P7, P10 & $7.46\times10^{-2}$ & $7.17\times10^{-2}$ & $10.63\times10^{-2}$ & $20.04\times10^{-2}$ \\
Trained on P4, P7, P10 & $7.52\times10^{-2}$ (-) & $7.62\times10^{-2}$ (-) & $5.37\times10^{-2}$ (+) & $21.20\times10^{-2}$ (-) \\
Trained on P1, P7, P10 & $8.85\times10^{-2}$ (-) & $7.63\times10^{-2}$ (-) & $6.94\times10^{-2}$ (+) & $32.95\times10^{-2}$ (-) \\
Trained on P1, P4, P10 & $8.94\times10^{-2}$ (-) & $7.14\times10^{-2}$ (+) & $12.90\times10^{-2}$ (-) & $28.87\times10^{-2}$ (-) \\
Trained on P1, P4, P7 & $8.86\times10^{-2}$ (-) & $5.78\times10^{-2}$ (+) & $10.06\times10^{-2}$ (+) & $24.10\times10^{-2}$ (-) \\
\bottomrule
\end{tabular}
\vspace{-0.48cm}
\end{table*}

\subsection{{PAD Efficiency Analysis}}

\subsubsection{Specific Settings}
{To demonstrate the superior cross-problem generalization efficiency of PAD’s single-model approach compared with other methods, we introduce the LLM-based solver OPRO \cite{deepmind2024LLMasOptimizer} and the traditional numerical solver Gurobi \cite{gurobi}, and conduct comparisons in terms of solution quality, algorithm construction cost, and runtime cost. Given that many existing works \cite{difusco,gdsg2024liang,liang2025diffsg} have already established the advantages of GDM-based solution generators over other discriminative models and DRL methods, and that PAD is specifically designed to explore cross-problem generalization, we do not include such baselines in this section. Furthermore, we extend Table \ref{tab_model_configs} by increasing the number of parallel samples in PAD, in order to showcase its solution quality and efficiency under different $p_{num}$ settings.}

\subsubsection{Results}
{As shown in Table \ref{tab_method_efficency}, while PAD does not match the optimal solver Gurobi in terms of solution quality and feasibility, it outperforms OPRO in most cases. This may partly stem from some problems lying beyond the knowledge scope of GPT-5. In terms of training or running efficiency, PAD is far more efficient and controllable in terms of hardware, computational resources, and time consumption. OPRO requires a massive number of model parameters and performs poorly on certain problems. Since GPT-5 is accessed via an API, its runtime hardware metrics are unavailable; however, its computational cost is expected to exceed that of PAD. Gurobi, on the other hand, relies heavily on manual coding and debugging, with solver construction typically requiring several hours per problem. Moreover, Gurobi is entirely incapable of handling problems outside its predefined design scope. In terms of inference efficiency, PAD substantially surpasses OPRO, whose model size and computational demands are disproportionate to the problem complexity, and it also consistently achieves lower latency than OPRO. Compared with Gurobi, PAD delivers comparable runtime performance while operating on common hardware platforms.}

{As shown in Fig. \ref{fig_parallel_sampling_exp}, we evaluate the impact of parallel sampling on solution quality under four configurations: $p_{num}=2,4,8,16$. Increasing $p_{num}$ significantly improves the final solution quality, approaching optimal performance on many problems at $p_{num}=16$, as this enables exploration of a broader distribution of high-quality solutions. In terms of efficiency, the four parallel sampling settings achieve average inference times of 117.25ms, 183.98ms, 315.31ms, and 600.47ms for single solution generation, respectively. These runtimes do not scale linearly with $p_{num}$ because the model’s nonlinear matrix operations prevent simple doubling of runtime with doubled sampling.}

\subsubsection{Summary}
{Overall, PAD demonstrates superior efficiency in model construction cost, inference cost, and inference latency compared to existing LLM-based solvers. Moreover, while its performance is already comparable to traditional numerical solvers in many aspects, PAD offers a more efficient and scalable potential advantage, particularly in cross-problem generalization.}

\subsection{{Analysis of Training Problem Selection}}

\subsubsection{Specific Settings}
{To analyze how the choice of training problems affects the model’s cross-problem generalization, we conduct four ablation studies by removing one problem at a time from the complete set of four training problems. In addition, this setup also serves an engineering purpose: to explore how clustering problems according to common mathematical properties influences the model’s within-cluster generalization performance.}

\subsubsection{Results}
{As shown in Table \ref{tab_training_problem_ablation}, we use the symbols (+) and (-) to indicate whether the performance of a given model is better or worse than that of the baseline model trained on the complete set of problems (the first row). From the diagonals of the last four rows, it is clear that the within-cluster generalization performance consistently degrades when the problem from that cluster is removed. This suggests that the excluded training problem indeed provides transferable knowledge for cross-problem generalization within its cluster, thereby supporting the validity of clustering problems according to common mathematical properties. Fundamentally, this also indicates that network optimization problems grouped by such mathematical properties share overlapping or transferable structures in the distribution of high-quality solutions.}

{For the clusters corresponding to the problems that were not removed, we argue that the observed performance gains or drops primarily result from the interplay between shared knowledge and competition for limited model capacity. In Table \ref{tab_training_problem_ablation}, the first and fourth columns both show degraded performance across all ablation models, indicating that problems in these clusters share substantial knowledge with those in other clusters (reflected at the neural network level by complementary gradient directions across problem). Removing training problems in these clusters reduces the availability of such shared knowledge, and the additional free model capacity is insufficient to compensate for the loss. Intuitively, both the first and fourth columns correspond to convex-like optimization problems, whose relatively simple high-quality solution distributions are less sensitive to model capacity than to shared knowledge. The second and third columns follow a similar pattern, but their performance is more strongly constrained by model capacity rather than shared knowledge. This is likely because the high-quality solution distributions of non-convex optimization, mixed-integer programming, and combinatorial optimization problems are inherently more diverse and complex.}

\subsubsection{Summary}
{The choice of training problems plays a critical role in determining the cross-problem generalization performance of the PAD model. Selecting representative problems from the same category based on common mathematical classifications can enhance the model’s ability to generalize to other problems within that category, while more complex problems also impose higher demands on model capacity. Moreover, problems grouped into different categories may still exhibit varying degrees of shared knowledge.}

\section{Conclusion and Future Directions}
{This paper has demonstrated a problem-aware learning approach for cross-problem network optimization, aiming to provide a promising new direction for equipping future foundation models in wireless networks with cross-problem solving capabilities. Specifically, PAD explicitly incorporates the mathematical representations of problem formulations via problem embedding, enabling the model to capture the mathematical structure of the optimization problems.}

{We have identified three key insights from experiments on ten representative network optimization problems, which vary in type and complexity. First, the mathematical expression–based feature embedding of network optimization problems contains rich structural and semantic information. When used as explicit conditional input, it enables a single model to achieve effective cross-problem generalization and generate high-quality solutions even on problems unseen during training.} Second, constructing an auxiliary constraint-aware module to identify infeasible solutions and incorporating it as a loss function can not only improve the feasibility of generated solutions but also enhance solution quality. {Third, clustering problems by their mathematical characteristics and selecting only representative problems from each cluster for training supports cross-problem generalization within each cluster. Furthermore, partial knowledge sharing across different clusters can also contribute to performance gains. Notably, with these three aspects realized, PAD surpasses LLM-based solvers in efficiency (such as algorithm construction and inference latency) while matching advanced numerical solvers, and simultaneously offers greater flexibility and scalability potential.}

{Additionally, several aspects of the current work warrant further investigation. First, the relationship between problem-feature embedding and solution generation. Although the current embedding strategy achieves a certain degree of cross-problem generalization, compared with tasks such as text-to-image generation or trajectory control, the linkage between training loss and solution quality through problem feature embedding remains relatively weak. Strengthening the model’s semantic understanding of problem feature embedding, as well as its direct influence on solution quality, is therefore essential. Second, improving the feasibility of generated solutions more efficiently remains a challenge. The current reliance on an external constraint-aware module and constraint loss still leaves room for improvement in terms of both efficiency and determinism. Third, optimizing model architecture and size is also necessary. Given the observed competition among multiple problems for model capacity, more compact yet effective neural designs are needed. Fourth, the classification and boundary of problems deserve further exploration. A deeper analysis of the transferability of high-quality solution distributions across objective functions is required to better clarify the limits of cross-problem generalization and to clarify how those boundaries may be expanded.}


\bibliographystyle{IEEEtran}  

{\appendix[Modeling of the Selected Optimization Problems]\label{sec_appendix}\small
We now present the detailed mathematical formulations and settings for the ten selected network optimization problems introduced in Section~\ref{sec_selected_problems}. The variable symbols used in each problem are isolated from those in other problems, with no reuse across problems.  
\begin{enumerate}
    \item \textbf{P1} aims to maximize the total transmission rate under constraints on total power and per-channel minimum rate requirements. Specifically, \( M \) denotes the number of channels, \( B \) [Hz] represents the default bandwidth of each OFDMA \cite{cao2024aiPROIEEE} channel, and \( P_{\text{total}} \) [W] denotes the total available power. The sets \( \{g_1, \ldots, g_M\} \) and \( \{N_1, \ldots, N_M\} \) represent the channel gains and noise power spectral density for each channel, respectively.\\
    Variable: Continuous power allocation $\{p_1,\ldots,p_M\}$[W] for each channels $m = \{1,\dots,M\}$. \\
    Objective: 
    $$\max_{\{p_1,\ldots,p_M\}} \sum_{m=1}^{M} B \log_{2}(1+\frac{g_m p_m}{B N_m})$$
    Constraints: 
    $$ \sum_{m=1}^{M} p_m \leq P_{\rm total}, $$
    $$B \log_{2}(1+\frac{g_m p_m}{B N_m}) \geq R_{\min}, $$
    $$ p_m\ge0, \forall m \in \{1,\ldots,M\}$$

    This modeling belongs to a classic category of wireless resource allocation problems \cite{userCentric2024debbah,tango2023infocom,gnn2023jsac}.

    \item \textbf{P2} aims to maximize the total transmission rate, under constraints on the total available spectrum resource block and minimum rate requirements for each channel. Specifically, \( M \) denotes the number of channels. Since in real-world scenarios, spectrum resources are typically allocated in discrete blocks rather than continuous values, we define $B_{\text{total}}$ as the total number of resource blocks and $b$ [Hz] as the bandwidth of each block. The sets \( \{g_1, \ldots, g_M\} \) and \( \{N_1, \ldots, N_M\} \) represent the channel gains and noise power spectral density for each channel, respectively. The \( \{p_1, \ldots, p_M\} \) represent the transmission powers for each channel. \\
    Variable: Discrete spectrum resource block allocation $\{B_1,\ldots,B_M\}$ for each channel $m = \{1,\dots,M\}$.\\
    Objective: $$\max_{\{B_1,\ldots,B_M\}} \sum_{m=1}^{M}B_m b\log_{2}(1+\frac{g_m p_m}{B_m b N_m})$$
    Constraints: $$\sum_{m=1}^M B_m\le B_{\rm total},$$
    $$ B_m b\log_{2}(1+\frac{g_m p_m}{B_m b N_m}) \geq R_{\min},$$
    $$ B_m\ge0, \forall m \in \{1,\ldots,M\}$$

    This represents another classic category of wireless resource allocation problems \cite{cao2024aiPROIEEE,subnetwork2023X,spectrumSharing2019}, where the inclusion of discrete optimization variables further enhances the realism of our problem set.

    \item \textbf{P3} aims to maximize the total transmission rate under constraints on total spectrum availability and fairness across channels based on predefined weights. Specifically, \( M \) denotes the number of channels. Given $B_{\text{total}}$ as the total number of resource blocks and $b$ [Hz] as the bandwidth of each block. The sets \( \{g_1, \ldots, g_M\} \) and \( \{N_1, \ldots, N_M\} \) represent the channel gains and noise power spectral density for each channel, respectively. The \( \{p_1, \ldots, p_M\} \) represent the transmission powers for each channel. Additionally, \( \{\alpha_1, \ldots, \alpha_M\} \) denotes the contribution factors associated with each connected terminal device, while $\theta$ represents the minimum transmission rate required by the system under a specified contribution ratio. \\
    Variable: Discrete spectrum resource block allocation $\{B_1,\ldots,B_M\}$ for each channel $m = \{1,\dots,M\}$.\\
    Objective:
    $$\max_{\{B_1, \ldots, B_M\}} \sum_{m=1}^{M} B_m b \log_{2} \left(1 + \frac{g_m p_m}{B_m b N_m}\right)$$
    Constraints:
    $$\sum_{m=1}^{M} B_m \leq B_{\text{total}},$$
    $$\frac{B_m b\log_{2} \left(1 + \frac{g_m p_m}{B_m b N_m}\right)}{\alpha_m} \geq \theta,$$
    $$ B_m\ge0, \forall m \in \{1,\ldots,M\}$$

    The problem’s constraint imposes different minimum rate requirements for each connection, which can be generalized to distributed weight synchronization tasks in typical federated learning settings \cite{survey2023federated}.

    \item \textbf{P4} aims to maximize the overall spectral efficiency, defined as the ratio of achievable rate to allocated bandwidth, under constraints on total spectrum and minimum per-channel rate requirements. Specifically, \( M \) denotes the number of channels. Given $B_{\text{total}}$ as the total number of resource blocks and $b$ [Hz] as the bandwidth of each block. The sets \( \{g_1, \ldots, g_M\} \) and \( \{N_1, \ldots, N_M\} \) represent the channel gains and noise power spectral density for each channel, respectively. The \( \{p_1, \ldots, p_M\} \) represent the channel transmission powers. \\
    Variable: Discrete spectrum resource block allocation $\{B_1,\ldots,B_M\}$ for each channel $m = \{1,\dots,M\}$.\\
    Objective: 
    $$\max_{\{B_1,\ldots,B_M\}} \sum_{m=1}^{M}\frac{\log_{2}(1+\frac{g_m p_m}{B_m b N_m})}{B_m}$$
    Constraints:
    $$\sum_{m=1}^M B_m\le B_{\rm total},$$
    $$B_m b\log_{2}(1+\frac{g_m p_m}{B_m b N_m}) \geq R_{\min}, $$
    $$B_m\ge0, \forall m\in\{1,\ldots,M\}$$

    Similarly, maximizing spectral efficiency is relevant in scenarios where spectrum conservation is necessary due to channel quality and terminal device power conditions \cite{cao2024aiPROIEEE,subnetwork2023X,spectrumSharing2019}.
    
    \item \textbf{P5} aims to maximize the power efficiency, defined as the ratio of transmission rate to consumed power, under constraints on total power and minimum rate requirements for each channel. Specifically, \( M \) denotes the number of channels, \( B \) [Hz] represents the default bandwidth of each OFDMA \cite{cao2024aiPROIEEE} channel, and \( P_{\text{total}} \) [W] denotes the total available power. The sets \( \{g_1, \ldots, g_M\} \) and \( \{N_1, \ldots, N_M\} \) represent the channel gains and noise power spectral density for each channel, respectively.\\
    Variable: Continuous power allocation $\{p_1,\ldots,p_M\}$ for each channels $m = \{1,\dots,M\}$. \\
    Objective: 
    $$\max_{\{p_1, \ldots, p_M\}} \quad \sum_{m=1}^{M}\frac{B \log_{2}\left(1 + \frac{g_m p_m}{B N_m}\right)}{p_m}$$ 
    Constraints: 
    $$ \sum_{m=1}^{M} p_m \leq P_{\text{total}},$$ 
    $$B \log_{2}\left(1 + \frac{g_m p_m}{B N_m}\right) \geq R_{\min},$$ 
    $$ \\ p_m\ge0, \forall m \in \{1,\ldots,M\}$$

    Maximizing power efficiency is particularly relevant in scenarios where power savings are required based on channel quality conditions \cite{userCentric2024debbah,tango2023infocom,gnn2023jsac}.

    \item {\textbf{P6} aims to optimize the refresh rate allocation of cached content at the edge server to minimize the expected hit latency. Each user $m \in \{1,\ldots,M\}$ generates requests following a Poisson process \cite{poisson2020cache,poisson2021refreshing}, with $\lambda_m$ [req/s] denoting the request arrival rate, $S_m$ [MB] the content size, and $\Delta_m$ [s] the freshness tolerance. The edge server has a total available bandwidth $\mu$ [MB/s], and refreshing incurs a unit cost $\alpha$ [s/MB]. An initial exponential formulation was simplified into a tractable fractional model for available solver support.}\\ 
    {Variable: Refresh allocation ratios $\{u_1,\ldots,u_M\}$, where $u_m \in [0,1]$ represents the fraction of bandwidth allocated to refreshing content $m$.}\\ 
    {Objective: 
    $$
    \min_{\{u_1,\ldots,u_M\}} 
    \ \omega_1 \sum_{m=1}^M \frac{\lambda_m S_m}{1+\frac{\Delta_m u_m \mu}{S_m}} 
    + \omega_2 \frac{L}{1-L} 
    + \alpha \mu L
    $$ 
    Constraints: 
    $$ L = \sum_{m=1}^M u_m, $$
    $$ \sum_{m=1}^M u_m \leq 1, $$
    $$ u_m \geq 0, \quad \forall m \in \{1,\ldots,M\} $$}
    
    {Refresh rate optimization is a crucial issue in edge server content caching scenarios \cite{refresh2019optimize}}.  

    \item {\textbf{P7} aims to maximize the total transmission rate by jointly allocating spectrum bandwidth and transmission power across multiple channels subject to minimum rate constraints. Specifically, $M$ denotes the number of channels, $B_{\text{total}}$ is the total number of resource blocks, and $b$ [Hz] is the bandwidth of each block. And $P_{\rm total}$ [W] is the total available transmission power. For each channel $m \in \{1,\ldots,M\}$, $g_m$ represents the channel gain and $N_m$ denotes the noise power spectral density.}\\ 
    {Variable: Discrete spectrum resource block allocation $\{B_1,\ldots,B_M\}$ and power allocation $\{p_1,\ldots,p_M\}$ [W].}\\ 
    {Objective: 
    $$
    \max_{\{B_1,\ldots,B_M,p_1,\ldots,p_M\}} 
    \sum_{m=1}^{M} B_m b \log_{2}\!\left(1+\frac{g_m p_m}{B_m b N_m}\right)
    $$ 
    Constraints: 
    $$ \sum_{m=1}^M B_m \leq B_{\rm total}, $$
    $$ \sum_{m=1}^M p_m \leq P_{\rm total}, $$
    $$ B_m \geq 0, \ p_m \geq 0, \quad \forall m \in \{1,\ldots,M\}, $$
    $$ B_m b \log_{2}\!\left(1+\frac{g_m p_m}{B_m b N_m}\right) \geq R_{\min}, \quad \forall m \in \{1,\ldots,M\} $$}
    
    {This modeling is more comprehensive and more difficult than considering spectrum or power resources alone, which is a classic mixed integer programming \cite{joint2021twc}.}

    \item {\textbf{P8} aims to minimize the expected hit latency by optimizing cache placement decisions at edge nodes. Specifically, $M$ denotes the number of candidate contents, $s_i$ [MB] represents the size of content $i$, $\lambda_i$ [req/s] is the request arrival rate for content $i$, $\alpha$ [s/MB] denotes the per-unit latency saving factor, and $C$ [MB] is the total cache capacity.} \\ 
    {Variable: Binary cache decision $\{x_i \in \{0,1\}\}$ indicating whether content $i$ is stored in the cache.} \\ 
    {Objective:
    $$
    \max_{\{x_i \in \{0,1\}\}} \sum_{i=1}^{M} \lambda_i \alpha s_i x_i
    $$
    Constraints:
    $$ \sum_{i=1}^{M} s_i x_i \leq C, $$
    $$ x_i \in \{0,1\}, \quad \forall i \in \{1,\ldots,M\}. $$}
    
    {This modeling is much simpler than refresh rate optimization and is a classic combinatorial optimization problem. \cite{caching2017jsac,content2018caching}.}

    \item {\textbf{P9} aims to minimize the overall task offloading cost through joint optimization of binary offloading decisions and continuous computational resource allocation. Specifically, $M$ denotes the number of users, $C^{\rm local}_m$ represents the local execution cost of task $m$, $C^{\rm trans}_m$ is the transmission cost, $C^{\rm offload}_m$ is the offloading computation cost. In practice, these costs are derived from parameters such as the task data size, computational workload requirements, and network conditions. However, in most solution approaches, the costs are first quantified based on the raw inputs and then used for optimization.} \\ 
    {Variable: Binary offloading decision $\{D_m \in \{0,1\}\}$ indicating whether task $m$ is offloaded, and continuous resource allocation $\{A_m \in [0,1]\}$.} \\ 
    {Objective:
    $$
    \min_{\{\substack{D_1,\ldots,D_M \\ A_1,\ldots,A_M}\}} \sum_{m=1}^{M} \big[(1-D_m)C^{\rm local}_m + D_m \big(C^{\rm trans}_m + \tfrac{C^{\rm offload}_m}{A_m}\big)\big]
    $$
    Constraints:
    $$ A_m \in [0,1], \quad \forall m \in \{1,\ldots,M\}, $$
    $$ D_m \in \{0,1\}, \quad \forall m \in \{1,\ldots,M\}, $$
    $$ \sum_{m=1}^{M} D_m A_m \leq 1. $$}
    
    {Computation offloading is a common network optimization problem in wireless network edge computing, which is a more difficult mixed integer programming problem \cite{yang2021MTFNN,xue2025jointtaskoffloadingresource}.}

    \item \textbf{P10} aims to minimize the maximum transmission delay across all channels, defined as the ratio between the data volume and the corresponding achievable transmission rate, under a constraint on the total transmit power. Specifically, \( M \) denotes the number of channels, \( B \) [Hz] represents the default bandwidth of each OFDMA \cite{cao2024aiPROIEEE} channel, and \( P_{\rm total} \) [W] denotes the total available power. And $\{d_1,\dots,d_M\}$ [bit] denotes the amount of data that needs to be transmitted for each connection. The sets \( \{g_1, \ldots, g_M\} \) and \( \{N_1, \ldots, N_M\} \) represent the channel gains and noise power spectral density for each channel, respectively.\\
    Variable: Continuous power allocation $\{p_1,\ldots,p_M\}$ for each channels $m = \{1,\dots,M\}$. \\
    Objective: 
    $$\min_{\{p_1, \ldots, p_M\}} \max_{m} \frac{d_m}{B \log_{2}(1 + \frac{g_m p_m}{B N_m})}$$
    Constraints: 
    $$\sum_{m=1}^{M} p_m \leq P_{\rm total},$$
    $$p_m\ge0, \forall m \in \{1,\ldots,M\}$$

    This type of problem represents a fundamental class of transmission tasks \cite{gdmRL2024jsac}. At the same time, as a maximization-oriented optimization problem, it also contributes to the diversity of our problem set.

\end{enumerate}

The selection of the above problems ensures that our problem set encompasses a diverse and realistic range of resource types, optimization variables, objectives, and constraints.

}



 

\begin{IEEEbiography}[{\includegraphics[width=1in,height=1.25in,clip,keepaspectratio]{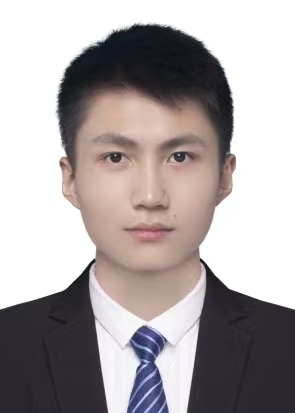}}]{Ruihuai Liang} received the B.E. degree in software engineering from Northwestern Polytechnical University (NPU), Xi'an, China, in 2023. He is currently pursuing the Ph.D. degree in computer science with Northwestern Polytechnical University, Xi'an, China. His current research interests focus on intelligent communication and networking.
\end{IEEEbiography}
\vspace{-0.4cm}

\begin{IEEEbiography}[{\includegraphics[width=1in,height=1.25in,clip,keepaspectratio]{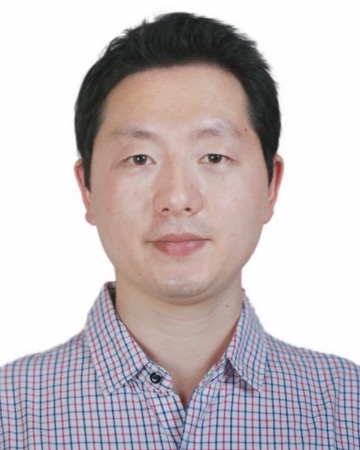}}]{Bo Yang} (Senior Member, IEEE) received the Ph.D. degree from Northwestern Polytechnical University (NPU) in 2017. From Oct. 2017 to Jan. 2020, he was a Postdoctoral Fellow with the Department of ECE, Prairie View A\&M University, TX, USA. From Feb. 2020 to April 2022, he was a Research Fellow with the EPD Pillar, Singapore University of Technology and Design (SUTD), Singapore. He is currently a Professor with the School of Computer Science, NPU. His research focuses on application of AI for wireless networking. Dr. Yang served as the Session Chair for IEEE Vehicular Technology Conference (VTC)2018-Fall, the Track Chair for EAI IoTaas 2020, and TPC Chair for IEEE PIMRC 2026. He also served as an Associate Editor for SN Computer Science. He received the CCF CWSN Best Paper Award in 2023 and IEEE VTC2025-Spring Best Student Paper Award.
\end{IEEEbiography}
\vspace{-0.4cm}

\begin{IEEEbiography}[{\includegraphics[width=1in,height=1.25in,clip,keepaspectratio]{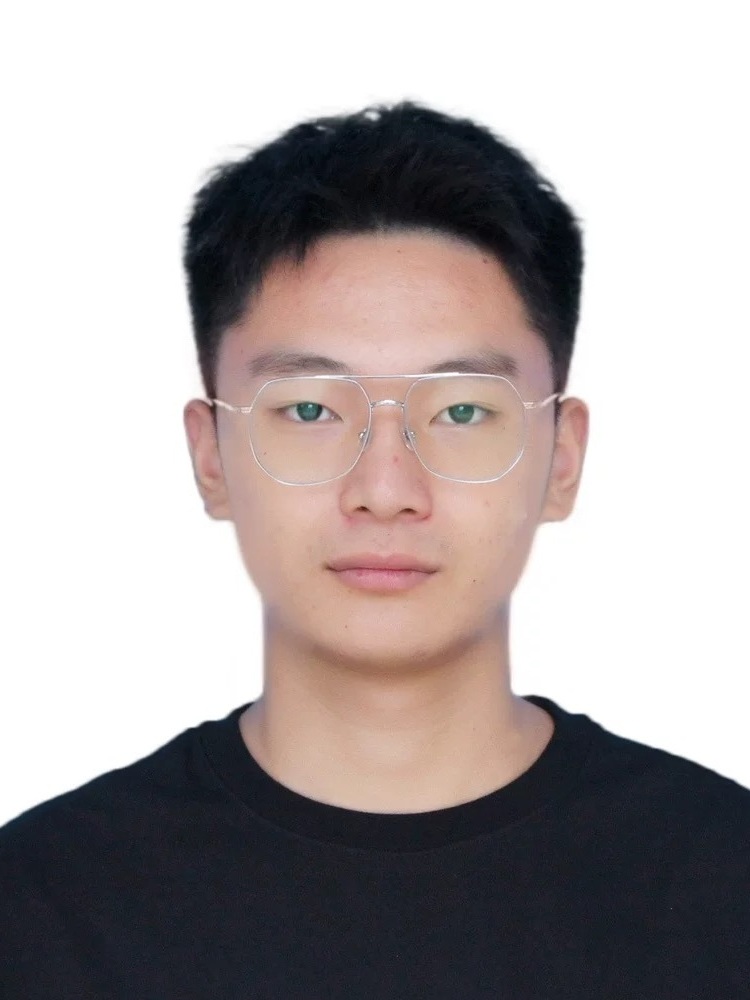}}]{Pengyu Chen} is an undergraduate student in Computer Science and Technology at Northwestern Polytechnical University (NPU), Xi'an, China, and is expected to obtain his B.E. degree in 2026. He will begin his Ph.D. studies at the College of Computer Science and Technology, Zhejiang University (ZJU), Hangzhou, China. His research interests include the theory of generative models and their applications.
\end{IEEEbiography}
\vspace{-0.4cm}

\begin{IEEEbiography}
[{\includegraphics[width=1in,height=1.25in,clip,keepaspectratio]{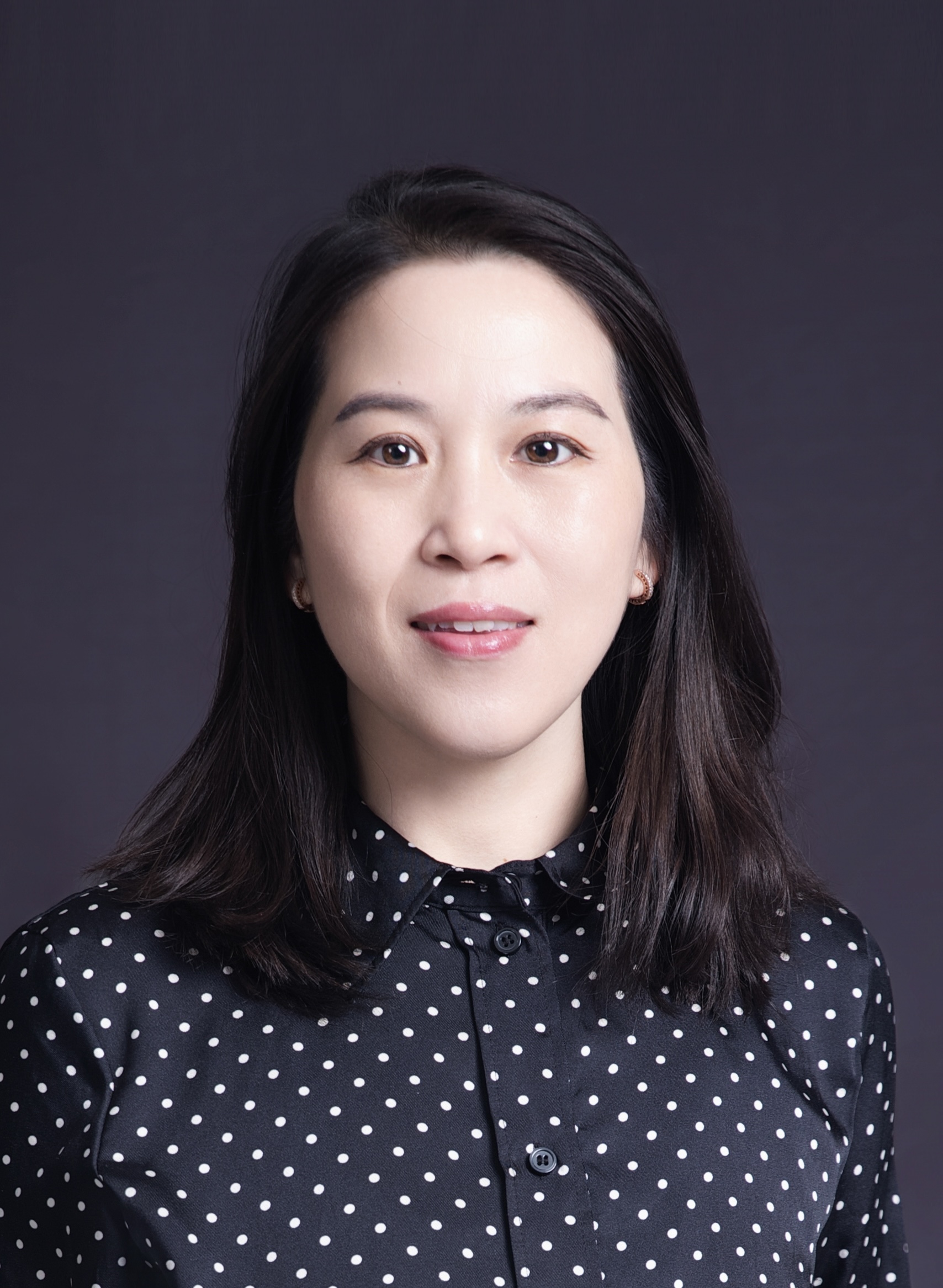}}]{Xuelin Cao} (Senior Member, IEEE) received her Ph.D. degree in communication engineering from Northwestern Polytechnical University, China, in 2019. From 2010 to 2012, she worked at Nokia Siemens Networks (NSN), Hangzhou, China, as an Engineer. From Sep. 2017 to Jan. 2020, she was a Visiting Scholar at the University of Houston, USA. From Feb. 2020 to Mar. 2023, she was a Research Fellow at Singapore University of Technology and Design, Singapore. She is currently a professor at Xidian University. Her research interests include medium access control (MAC) protocols, 5G/6G, reconfigurable intelligent surface (RIS), multiaccess edge computing (MEC), and artificial intelligence (AI). She was the TPC member of ICC, GLOBECOM, WCNC, VTC and WCSP. She served as the Session Chair for GLOBECOM 2023 and VTC 2024 Spring. She is an Associate Editor of SN Computer Science.
\end{IEEEbiography}
\vspace{-0.4cm}

\begin{IEEEbiography}[{\includegraphics[width=1in,height=1.25in,clip,keepaspectratio]{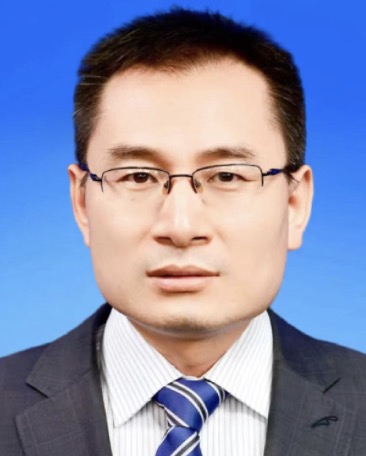}}]{Zhiwen Yu} (Senior Member, IEEE) received the PhD degree in computer science from Northwestern Polytechnical University, Xi’an, China, in 2005. He is currently the vice president of Harbin Engineering University, Harbin, China, and a professor with Northwestern Polytechnical University. He was an Alexander Von Humboldt fellow with Mannheim University, Germany, and a research fellow with Kyoto University, Kyoto, Japan. His research interests include ubiquitous computing, mobile crowd sensing, and human computer interaction.
\end{IEEEbiography}
\vspace{-0.4cm}

\begin{IEEEbiography}[{\includegraphics[width=1in,height=1.25in,clip,keepaspectratio]{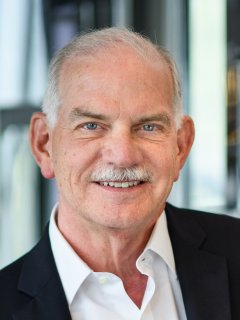}}]{H. Vincent Poor} (S’72, M’77, SM’82, F’87) received the Ph.D. degree in EECS from Princeton University in 1977.  From 1977 until 1990, he was on the faculty of the University of Illinois at Urbana-Champaign. Since 1990 he has been on the faculty at Princeton, where he is currently the Michael Henry Strater University Professor. During 2006 to 2016, he served as the dean of Princeton’s School of Engineering and Applied Science, and he has also held visiting appointments at several other universities, including most recently at Berkeley and Caltech. His research interests are in the areas of information theory, stochastic analysis and machine learning, and their applications in wireless networks, energy systems and related fields. Among his publications in these areas is the book Machine Learning and Wireless Communications.  (Cambridge University Press, 2022). Dr. Poor is a member of the National Academy of Engineering and the National Academy of Sciences and is a foreign member of the Royal Society and other national and international academies. He received the IEEE Alexander Graham Bell Medal in 2017.
\end{IEEEbiography}
\vspace{-0.4cm}

\begin{IEEEbiography}[{\includegraphics[width=1in,height=1.25in,clip,keepaspectratio]{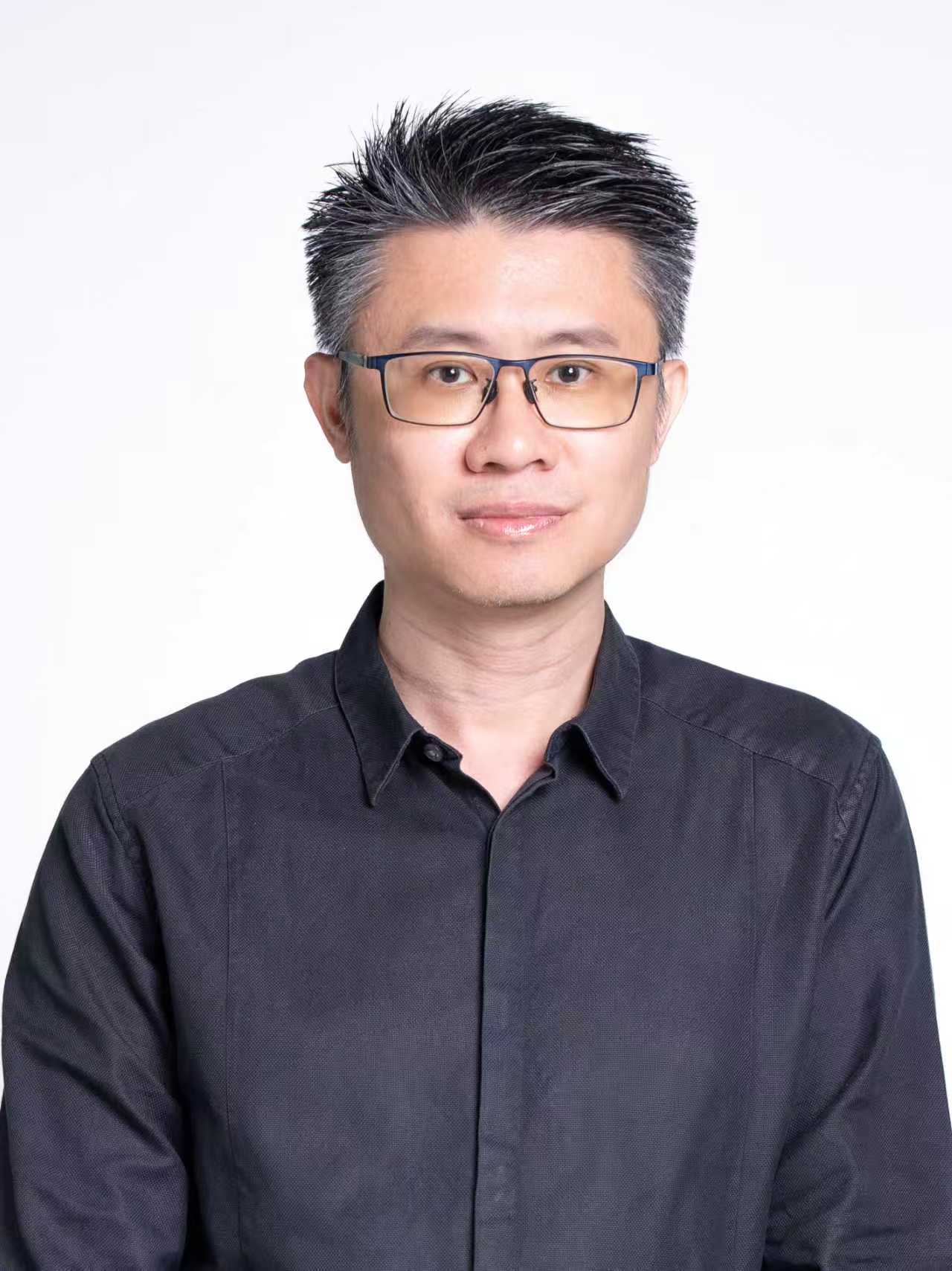}}]{Chau Yuen} (S'02-M'06-SM'12-F'21) received the B.Eng. and Ph.D. degrees from Nanyang Technological University, Singapore, in 2000 and 2004, respectively. He was a Post-Doctoral Fellow with Lucent Technologies Bell Labs, Murray Hill, in 2005. From 2006 to 2010, he was with the Institute for Infocomm Research, Singapore. From 2010 to 2023, he was with the Engineering Product Development Pillar, Singapore University of Technology and Design. Since 2023, he has been with the School of Electrical and Electronic Engineering, Nanyang Technological University, currently he is Provost’s Chair in Wireless Communications, Assistant Dean in Graduate College, and Cluster Director for Sustainable Built Environment at ER@IN. Dr Yuen currently serves as an Editor-in-Chief for Springer Nature Computer Science. He is listed as Top 2\% Scientists by Stanford University, and also a Highly Cited Researcher by Clarivate Web of Science from 2022. 

\end{IEEEbiography}



\vfill

\end{document}